\documentclass[aps,prl,twocolumn,superscriptaddress,longbibliography]{revtex4-2}

\usepackage{natbib}

\usepackage{amsmath,amssymb,graphicx,mathtools,dsfont,bbold}
\usepackage[dvipsnames]{xcolor}
\usepackage[T1]{fontenc}
\usepackage{hyperref}
\usepackage{soul}
\hypersetup{citecolor=red,colorlinks=true,urlcolor=blue}

\newcommand{\bk}{\mathbf{k}}
\newcommand{\bq}{\mathbf{q}}
\newcommand{\ve}{\varepsilon}

\newcommand{\att}{\mathrm{a}}
\newcommand{\rep}{\mathrm{r}}

\newcommand{\NCU}{Institute of Physics, Faculty of Physics, Astronomy and Informatics, Nicolaus Copernicus University
in Toru\'n, Grudzi\c{a}dzka 5, 87-100 Toru\'n, Poland }
\newcommand{\PKS}{Max-Planck-Institut f\"ur Physik komplexer Systeme, N\"othnitzer Str. 38, 01187 Dresden, Germany}
\newcommand{\BEC}{INO-CNR  BEC  Center and Dipartimento di  Fisica,  Universit\`a di Trento, 38123  Trento, Italy}
\newcommand{\UK}{Institut f\"ur Theoretische Physik, Universit\"at zu K\"oln, Z\"ulpicher Stra{\ss}e 77, 50937 Cologne, Germany}
\newcommand{\TIFPA}{Trento Institute for Fundamental Physics and Applications, INFN, 38123, Trento, Italy}
\newcommand{\AUG}{Theoretical Physics III, Center for Electronic Correlations and Magnetism,
Institute of Physics, University of Augsburg, 86135 Augsburg, Germany}

\begin{document}
\author{Tomasz Wasak}\email{twasak@umk.pl}\affiliation{\NCU}\affiliation{\PKS}  
\author{Matteo Sighinolfi}\affiliation{\BEC}
\author{Johannes Lang}\affiliation{\PKS}\affiliation{\UK}
\author{Francesco Piazza}\email{piazza@pks.mpg.de}\affiliation{\AUG}\affiliation{\PKS}  
\author{Alessio Recati}\email{alessio.recati@ino.cnr.it}\affiliation{\BEC}\affiliation{\TIFPA}

\title{Decoherence and momentum relaxation in Fermi-polaron Rabi dynamics:\\ a kinetic equation approach}
\date{\today}

\begin{abstract}
Despite the paradigmatic nature of the Fermi-polaron model, the theoretical description of its nonlinear dynamics poses challenges. Here, we apply a quantum kinetic theory of driven polarons to recent experiments with ultracold atoms, where Rabi oscillations between a Fermi-polaron state and a non-interacting level were reported. The resulting equations separate decoherence from momentum relaxation, with the corresponding rates showing a different dependence on microscopic scattering processes and quasi-particle properties. We describe both the polaron ground state and the excited repulsive-polaron state and we find a good quantitative agreement between our predictions and the available experimental data without any fitting parameter.
Our approach not only takes into account collisional phenomena, but also it can be used to study the different roles played by decoherence and the collisional integral in the strongly interacting highly-imbalanced mixture of Fermi gases. 
\end{abstract}
\maketitle


\textit{Introduction.}--- Polarons, i.e., impurities dressed by its environment, plays a paradigmatic role in the understanding of many-body properties in a variety of physical systems. The Fermi polaron - a particle interacting with a reservoir of free fermions - became the object of intense study after its experimental realization with ultracold atomic gases \cite{zwierlein2009polaron,nascimbene2009pol,koschorreck2012attractive,massignan2014polarons,cetina2016ultrafast,gal2020polmol}, and more recently in monolayer semiconductors \cite{imamoglu2020interacting,emmanuele2020highly}. Studying this problem has offered valuable insights into the many-body physics of Fermi-Fermi or Bose-Fermi mixtures \cite{massignan2014polarons}.

The high level of control allows the investigation of the dynamics of Fermi polarons  \cite{kohstall2012metastability}. In particular, experiments have studied the fate of coherent Rabi dynamics between a non-interacting impurity state and a polaron state \cite{scazza2017repulsive,oppong2019,haydn2020repulsive}, where collisional induced relaxation and decoherence are at play. 

In thermal equilibrium, due to the simplicity of the Fermi-polaron model, fully solvable microscopic approaches, like simple variational Ans\"atze \cite{chevy2006universal,combescot2007normal}, have produced quantitatively valid predictions. However, the theoretical description of the problem becomes more challenging when considering the dynamics. For instance, the theoretical description of the Fermi-polaron Rabi dynamics based on a time-dependent variational approach revealed some nontrivial features \cite{RabiPolaro-Jesper2016}, like the absence of decay from the repulsive to the attractive branch, which were required to achieve agreement with the experimental data \cite{haydn2020repulsive}.

Moreover, for such a problem -- related to the open quantum system problem of a spin in a bath -- it would be desirable to have a density matrix description and to identify the role of the various relaxation mechanism. 

In this work, we apply a quantum kinetic approach \cite{wasak2021fermi} to study the Fermi-polaron Rabi dynamics, which allows us to describe the system in terms of the evolution of the density matrix of a two-level system, whose properties are dressed by many-body effects. Importantly, we can distinguish between decoherence and momentum relaxation, whose rates show a different dependence on scattering processes and quasi-particle properties. 
Within a simple approximation for the scattering between the impurity and the bath, the solution of our quantum kinetic equations is shown to be in good agreement with the available experimental data without any fitting parameters. This indicates that the experiments realize the situation where polarons (and not the bare particles) perform Rabi oscillations.

\begin{figure}[t]
  \includegraphics[width=\columnwidth]{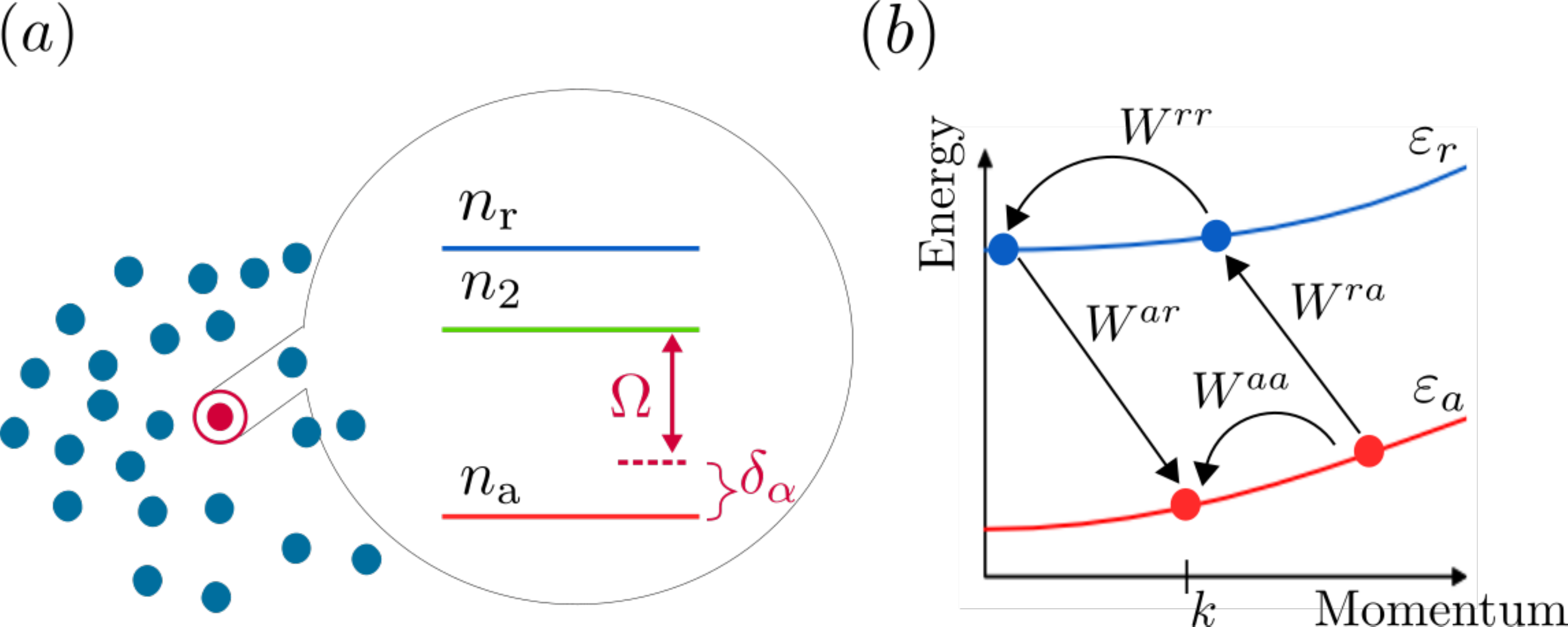}
\caption{{\bf(a)} A highly imbalanced mixture of atoms in state $|1\rangle$ (majority, blue) and $|2\rangle$ (minority, red dot in a circle) is held at temperature $T$ with zero interspecies interaction. The Rabi coupling $\Omega$ drives the transitions between states $|2\rangle$ and $|3\rangle$. In the latter, minority atoms occupy repulsive and attractive Fermi-polaron branches, formed due to the interaction with the majority component, while $\delta_\alpha$ is the detuning between the non-interacting state and the polaron levels (here only $\alpha=\att$ is shown). 
{\bf (b)} Polaron dispersion relations for atoms in state $|3\rangle$. The interaction with the majority atoms induces inter- and intra-band transitions, with rates $W^{\alpha\beta}$, where $\alpha,\beta\in\{\att,\rep\}$. 
}\label{fig1}
\end{figure}

\textit{Model and kinetic equations.}---We consider a homogeneous system composed of a bath of atoms in state $|1\rangle$, and impurity atoms in state $|3\rangle$, which are Rabi-coupled to a non-interacting state $|2\rangle$ (Fig.~\ref{fig1} ($a$)). All the states are different internal levels of the same isotope, and, therefore, they have the same mass $m$. The Hamiltonian of the system reads
\begin{equation}\label{eq:totHam}
    \hat H = \sum_{i=1,2,3}\hat H_i + \hat H_\mathrm{int} + \hat H_\mathrm{\Omega},
\end{equation}
where $\hat H_i = \sum_\bk (\bk^2/2m)\hat c_{\bk,i}^\dagger \hat c_{\bk,i}$, with $\hat c_{\bk,i}^{(\dagger)}$ the annihilation (creation) operators of a Fermi atom in the state $|i\rangle$ with momentum $\bk$ (hereafter $\hbar=1$) and  $\hat H_\mathrm{int} = \frac{U_0}{V} \sum_{\bk,\bk',\bq} \hat c_{\bk'-\bq,1}^\dagger \hat c_{\bk+\bq,3}^\dagger \hat c_{\bk,3}\hat c_{\bk',1}$ is the interaction potential.  In absence of the Rabi coupling, when the impurity is in the state $|3\rangle$, the interaction with the bath leads to the emergence of quasi-particles known as Fermi polarons. In addition to a negative energy attractive branch, $\att$, there exists a meta-stable repulsive, $\rep$, polaron branch at positive energy. The polaron dispersion relations at small momenta are given by $\ve_\alpha(\bk)\approx 
E_\alpha + \bk^2/2m^*_\alpha$ ($\alpha=\att$ or $\rep$), where $E_\alpha$ -- with $E_\att<0$ and $E_\rep>0$ -- and $m^*_\alpha>m$ are referred to as the polaron energy and polaron effective mass, respectively. The branches are also characterised by the quasi-particle weight $0<Z_\alpha(\bk)\leqslant1$.
The last term in Eq.~\eqref{eq:totHam} induces the Rabi oscillations between the states $|2\rangle$ and $|3\rangle$, and, in the rotating wave approximation, can be written as
\begin{equation}\label{eq:Ham}
    \hat H_\Omega =  \sum_\bk \bigg[ \frac{\Omega}{2}  ( \hat c_{\bk,3}^\dagger \hat c_{\bk,2} + \hat c_{\bk,2}^\dagger \hat c_{\bk,3} ) + \Delta \hat c_{\bk,2}^\dagger\hat c_{\bk,2}\bigg],
\end{equation}
where $\Omega$ is the bare Rabi frequency and $\Delta$ is the bare detuning from the transition $|2\rangle\leftrightarrow|3\rangle$. 

In the experiments by Scazza {\sl et al.}~\cite{scazza2017repulsive}
for equal masses and by Kohstall {\sl et al.}~\cite{kohstall2012metastability} for large mass imbalance, as well as by Oppong {\sl et al.}~\cite{oppong2019} in quasi-2D geometry, it has been shown that it is possible to drive long-lived coherent Rabi oscillations between the non-interacting state and both the repulsive or the attractive polaron states. The resonant energies and the renormalized Rabi frequencies of the oscillations were found to be in reasonable agreement with an analysis based
on the assumption that $\Omega$ does not affect the polaron properties. 
On the other hand, the decay rate of the oscillations, especially for the supposedly long-lived attractive polaron, has not found a proper explanation yet.  Recently, a variational approach has been able to capture the dynamics for the repulsive branch~\cite{haydn2020repulsive}.

Our aim is to provide an equation of motion for the impurity, which is able to take into account the quasi-particle nature of the polarons, and to explain how the static polaron properties modify the Rabi oscillations. The main result of our study is the set of equations for the single-particle density matrix:
\begin{subequations}\label{eq:keq}
\begin{eqnarray}
&&\dot n_\alpha - i \frac{Z_\alpha \Omega}{2} (f_{2\alpha} - f_{2\alpha}^*) = I_{\alpha}, \label{keq_pol}\\
&&\dot n_2 + i \frac{\Omega}{2} (f_{2\alpha} - f_{2\alpha}^*) = 0, \label{keq_n2} \\
&&\dot f_{2\alpha} - i  \tilde Z_\alpha \delta_\alpha f_{2\alpha} \!+\! i \frac{\tilde Z_\alpha \Omega}{2} (n_2-n_\alpha)
\!=\!- \frac{\Gamma_\alpha^\mathrm{dec}}{2} f_{2\alpha}, \label{keq_f}
\end{eqnarray}
\end{subequations}
where we dropped the time $t$ and $\bk$ for brevity, and {$\tilde Z_\alpha = 2/(1+1/Z_\alpha)$}. The effective detuning $\delta_\alpha(\bk)=\ve_\alpha(\bk)-\ve_2(\bk)-\Delta\simeq 0$ determines whether the attractive $\alpha=\att$ or the repulsive $\alpha=\rep$ polaron is involved in the dynamics. In this notation, $n_\alpha(\bk,t)$ is the occupation of the polaron branch $\alpha$ at momentum $\bk$, $n_2(\bk,t)$ is the occupation of the state $|2\rangle$, which is coherently coupled to the branch $\alpha$, and $f_{2\alpha}(\bk,t)$ is the coherence between atoms in the state $|2\rangle$ and $\alpha$ polarons.

The quantum kinetic equations~\eqref{eq:keq} can be obtained {following the general approach established by Kadanoff and Baym~\cite{kadanoff2018quantum} for time-dependent Green's functions, thereby} extending previous works (see, e.g., \cite{ruckenstein89, kamenev2011field}) on the derivation of the kinetic equations for spin-$1/2$ Fermi quantum fluids in magnetic fields \footnote{our Rabi coupling term can be seen as an effective magnetic field ${\bf H}=(\Omega,0,\Delta)$} to three level systems in the highly imbalanced case, i.e., in the impurity limit (see the Supplemental Information (SM) for more details~\cite{sm}).
Here, we briefly discuss the main approximations. Within the usual Kramers-Moyal \cite{kamenev2011field} expansion, necessary to derive time-local equations, we drop the back-flow term \cite{Ivanov2001} (see also below) and assume that the polaron spectral properties, in particular the parameters $Z_\alpha$, $E_\alpha$ and $m_\alpha$, are time-independent. Moreover, the equations are derived by projecting on the energy shell of the impurity interacting with the equilibrium bath (see, e.g., \cite{combescot2007normal}), i.e., the polaron branch unmodified by the driving laser. 

The left-hand side of the kinetic equations in Eq.~(\ref{eq:keq}) predicts coherent oscillations with a renormalized Rabi frequency  {$\sqrt{Z_\alpha\Omega^2 + \tilde Z_\alpha^2 \delta_\alpha^2}$}. For {$\delta_\alpha = 0$, we obtain $\sqrt{Z_\alpha}\Omega$ which is} in agreement with the expression obtained by using a stationary variational Ansatz, which already received experimental verification ~\cite{kohstall2012metastability,scazza2017repulsive}. There are instead no measurement of the role of the detuning in determining the polaron oscillation, which, according to our result, is non-trivial. Interestingly a very recent experiment~\cite{nir2023strongly} has investigated such dependence for large $\Omega$ paving the way to further analysis on the role of the dressing in the polaron dynamics. 

The right-hand side of Eq. (\ref{eq:keq}), due to collisions between minority and majority atoms, contains i) the redistribution of the polaron population $n_\alpha$, described by the collision integral $I_\alpha$, and ii) the loss of coherence between the $\alpha$ polaron and the non-interacting state, described by the decoherence rate $\Gamma^{\mathrm{dec}}_{\alpha}$.
The population of the $\alpha$ polaron branch can be changed as a result of both inter- and intra-branch collisions.
In the impurity limit $n_\alpha(\bk)\approx0$, the decoherence rate reads
\begin{equation}
\label{decoh}
    \Gamma^{\mathrm{dec}}_{\alpha}(\bk) \!=\!
    \frac{2}{1+Z_\alpha(\bk)}\! \frac{1}{V}\!\!
    \sum_{\bk',\beta}W_{\bk'\bk}^{\beta\alpha},
\end{equation}
and the collision integral
\footnote{Beyond the impurity limit Eq. (\ref{I}) would also contain terms with $f_{2\alpha}$, however, they are not relevant for the physics discussed in the present Letter.} 
 takes the intuitive form $I_\alpha= \sum_\beta I_{\alpha\beta}$, where
\begin{equation}
\label{I}
    I_{\alpha\beta}(\bk) = \frac{1}{V} \sum_{\bk'}\bigg[
    W_{\bk\bk'}^{\alpha\beta} n_\beta(\bk') 
    - W_{\bk'\bk}^{\beta\alpha} n_\alpha(\bk) 
    \bigg],
\end{equation}
where $W_{\bk'\bk}^{\beta\alpha}$ is the transition rates  from the $\alpha$-branch with momentum $\bk$ to the $\beta$-branch with momentum $\bk'$.

The above expressions show that the redistribution of population and the decoherence, originating from collisions, are different in nature. The redistribution results from the imbalance between \emph{in} and \emph{out} scattering processes (see  Eq.~\eqref{I}), the density independent decoherence rate, Eq.~\eqref{decoh}, is due to possible scattering processes {between} a polaron $\alpha$ at momentum $\bk$ and a polaron $\beta$ at momentum $\bk'$.
The population redistribution drives the minority atoms towards thermal equilibrium with the majority ones, being indeed $I_{\alpha\beta}=0$ considering the Boltzmann equilibrium distribution $n_\alpha^{\mathrm{eq}}(\bk) \propto e^{-\ve_\alpha(\bk)/T}$, with $T$ the temperature of the majority component.
As expected, notice that for decreasing quasi-particle weight the pre-factor $2/(1+Z_\alpha)$ in Eq.~\eqref{decoh} increases the role of the decoherence term compared to the redistribution rate.

The transition rates are due to the scattering of a polaron with an atom of the majority component. They are given by the Fermi's golden rule 
\begin{eqnarray}
    W_{\bk,\bk'}^{\alpha\beta} \!\!&=&\!\! \frac{2\pi}{V}\!\! \sum_{\mathbf{Q}}
        |T_\mathrm{sc}(\mathbf{Q}, \ve_\beta(\bk')\!+\!\ve_1(\mathbf{Q}-\bk'))|^2
        Z_\alpha(\bk) Z_\beta(\bk') \nonumber \\
        &&\times \delta( \ve_\alpha(\bk) + \ve_1(\mathbf{Q}-\bk) 
            - \ve_1(\mathbf{Q}-\bk') - \ve_\beta(\bk')) \nonumber \\
        && \times n_1^\mathrm{eq}(\mathbf{Q}-\bk') 
            [ 1 -  n_1^\mathrm{eq}(\mathbf{Q}-\bk) ],
            \label{W}
\end{eqnarray}
where $T_\mathrm{sc}(\mathbf{Q}, \varepsilon)$ is the scattering matrix with $\mathbf{Q}$ and $\varepsilon$ the total momentum and the total energy of the particles entering the collision,
$n_1^\mathrm{eq}(\bk) = 1/[e^{(\ve_1(\bk) - \mu)/T} + 1]$ is the thermal distribution of the majority atoms with chemical potential $\mu$, where $\ve_1(\bk) = \bk^2/2m$. Notice how the quasi-particle weights of the initial and final states renormalize the transition rates.

For the attractive polaron branch, the decoherence rate $\Gamma_\att^\mathrm{dec}(\bk)$ is particularly simple at small momenta, since the scattering of the majority atoms takes place in the vicinity of the Fermi surface. We may write it as
\begin{eqnarray}
&&\Gamma_\att^\mathrm{dec}(k) \approx \frac{\tilde Z_\att(0) Z_\att(0) m^2 |T_\mathrm{sc}(k_F, \ve_F \!+\!\ve_\att(k))|^2}{4\pi^3} \times \nonumber\\ 
    &&
    \int_0^\infty k'^2 dk' \frac{\ve_\att(k')\!-\!\ve_\att(k)+T \ln\!\bigg[\frac{1+e^{\frac{\ve_F}{T}}}{e^{\frac{\ve_F}{T}}+e^{\frac{\ve_\att(k')\!-\!\ve_\att(k)}{T}}}\bigg]}{\max(k',k) [e^{\frac{\ve_\att(k')\!-\!\ve_\att(k)}{T}}\!-\!1]}, 
\label{decoh0}
\end{eqnarray}
and therefore the timescale of the decoherence rate in the Rabi oscillations is set by $|T_\mathrm{sc}(k_F, \ve_F + \ve_\att(0))|^2$ and a $k$-dependent function originating from the phase space, where $\ve_F=k_F^2/(2m)$ is the Fermi energy and $k_F$ is the Fermi wave vector.

\textit{Setting and parameters.}---
In the following, we compare the solution of our kinetic equations to the experimental results from Ref.~\cite{scazza2017repulsive}. 
The impurity atoms are initially prepared in the non-interacting state $|2\rangle$ and in thermal equilibrium with the bath at temperature $T$. The initial condition for solving the kinetic equations are thus $n_\alpha=0$, $f_{2\alpha}=0$ and $n_2^\mathrm{eq}(\bk) = [e^{(\ve_2(\bk) - \mu_2)/T)} + 1]^{-1}$,
where the chemical potential $\mu_2$ fixes the imbalance $x\equiv\rho_2/\rho_1$, where $\rho_i = (1/V)\sum_\bk n_i^\mathrm{eq}(\bk)$. 
We {compute} the polaron parameters $Z_\alpha(\bk)$ \footnote{For the sake of simplicity, given its very weak momentum dependence, we replace $Z_\alpha(\bk)$ with $Z_\alpha(0)$}, $E_\alpha$, $m_\alpha$, and $T_\mathrm{sc}(\bk,\omega)$ within the
non-self-consistent $T$-matrix approach \cite{combescot2007normal}.
In particular, the scattering matrix reads
\begin{align}
     T_\mathrm{sc}^{-1}(\mathbf{Q},\omega)\!=\!\frac{m}{4\pi a}\!-\!\frac{1}{V}\!\!\sum_\mathbf{k}\!\left(\!\frac{1-n_1^\text{eq}(\mathbf{k}+\mathbf{Q}/2)}{\omega-\frac{\mathbf{Q}^2+4\mathbf{k}^2}{4m}+i 0^{+}}\!+\!\frac{m}{\mathbf{k}^2}\!\right),
\end{align}
where $a$ is the $s$-wave scattering length  between the atoms in states $|1\rangle$ and  $|3\rangle$, whose relation to the two-body contact potential reads: 
$U_0^{-1}=m/(4\pi a)-(1/V)\sum_\mathbf{k}m/k^2$. 
This approach, even close to unitarity, compares reasonably well with experimental outcomes and quantum Monte Carlo calculations for the polaron's energy, mass and residue~\cite{scazza2017repulsive,pilati2010itinerant}. 

To be consistent with the experiment~\cite{scazza2017repulsive}, we set
$T=0.135\ve_F$, $\Omega = 0.7 \ve_F$ and the {imbalance} $x=0.15$.  We take $k_F$ equal to the effective $\kappa_F$ from Ref.~\cite{scazza2017repulsive}. We numerically determine the observable $N_2/(N_2+N_3)$, where $N_i$ is the total atom number in the state $i=2,\,3$, and $N_2(t)+N_3(t)=N_2(t=0)$.

\begin{figure}[t]
  \centering
  \includegraphics[width=1.0\columnwidth]{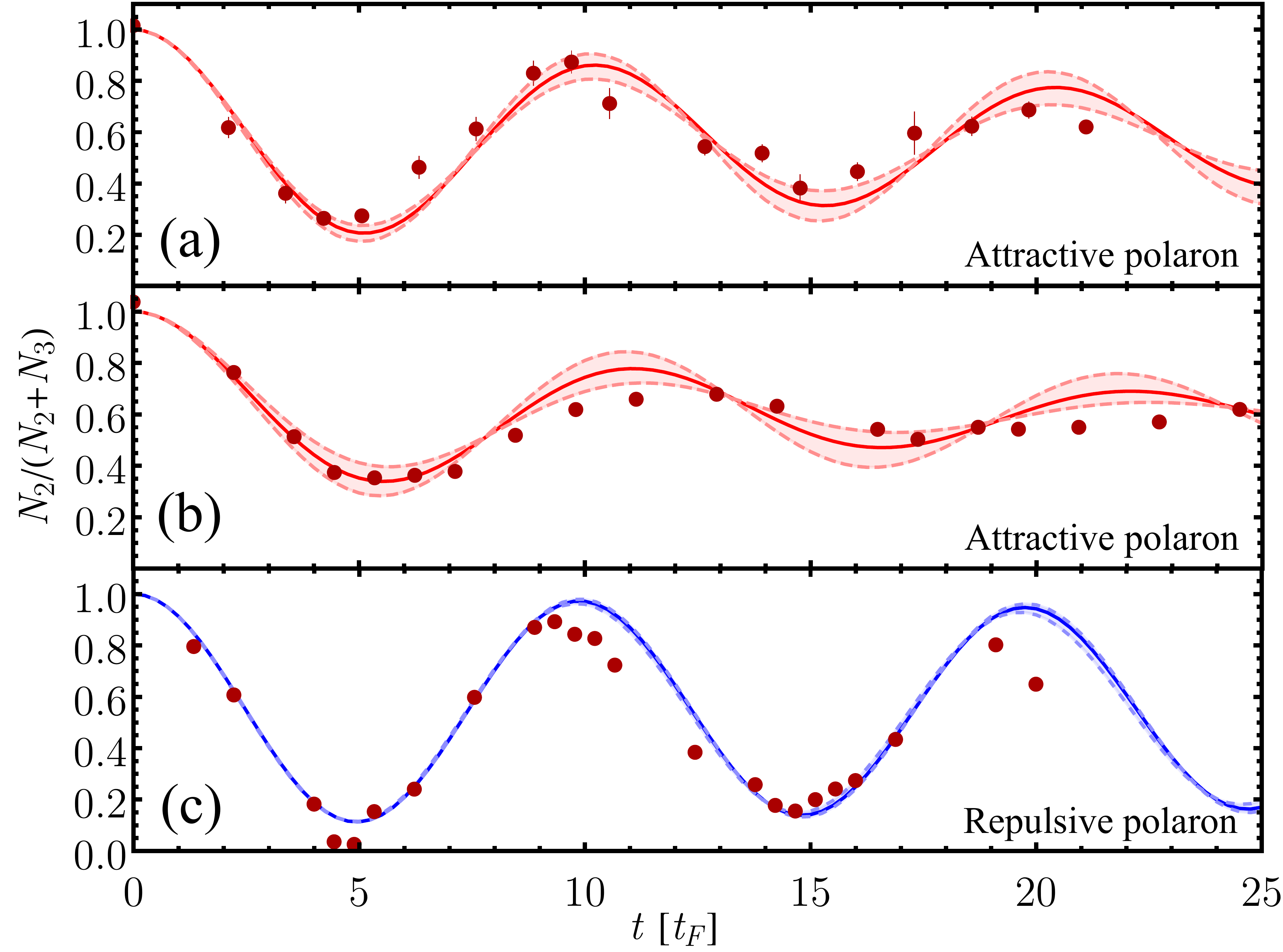}
\caption{Rabi oscillations of the attractive polarons for $1/(k_Fa)=$0 (a) and 0.25 (b) as a function of time $t$ (in units of $t_F=1/\ve_F$) and of the repulsive polaron
 for $1/(k_Fa) = 1.27$~(c). The red dots are from the experiment~\cite{scazza2017repulsive}. The solid red line is from Eq.~\eqref{eq:keq} and the non-self-consistent $T$-matrix approximation in Eq.~\eqref{W}. The red shaded region shows the confidence interval for $20\%$ relative uncertainty in the temperature. The parameters: $\Omega=0.7\ve_F$, $T=0.135\ve_F$.
}
\label{fig_att}
\end{figure}

\textit{{Rabi oscillations}.}---
To describe Rabi oscillations, we set $\delta_\alpha|_{\bk=0}\!\!=\!\!0$.
In Fig.~\ref{fig_att}a and~\ref{fig_att}b, we show the dynamics for  $1/(k_Fa)=0,\, 0.25$, for which the polaron parameters $(E_\att/\varepsilon_F,\,  m_*/m,\, Z_\att(0))$ entering in the simulation are $( -0.625,\, 1.16,\, 0.775)$ and $( -0.858,\, 1.29,\, 0.673)$, respectively. The repulsive polarons are considered to not be populated, due to small $Z_\rep$ and significant detuning. 
The shaded region takes into account the experimental uncertainty in the determination of the temperature, i.e., $T\pm \Delta T=0.135\ve_F (1\pm 20\%) $. The agreement with the experimental data is quite remarkable. While the renormalised oscillation frequency $ \sqrt{Z_\att}\Omega$ can be explained just by static calculations, the decay rate within more standard variational approaches is usually found to be too small \cite{RabiPolaro-Jesper2016} or ad-hoc assumption have to be done \cite{massignan2014polarons}. Not only our approach is built to properly take into account collisional phenomena, but we disentangle the role played by decoherence and by the collisional integral. In particular, we find that for the Rabi oscillation dynamics, neglecting the collisional integral is still a good approximation.
For completeness in Fig.~\ref{fig_att}c, we show the comparison with the repulsive polaron experimental data \cite{scazza2017repulsive} at $1/(k_F a)=1.27$, where however the collisions are playing a minor role (see SM for further discussion).


\begin{figure}[t]
  \centering
  \includegraphics[width=1.0\columnwidth]{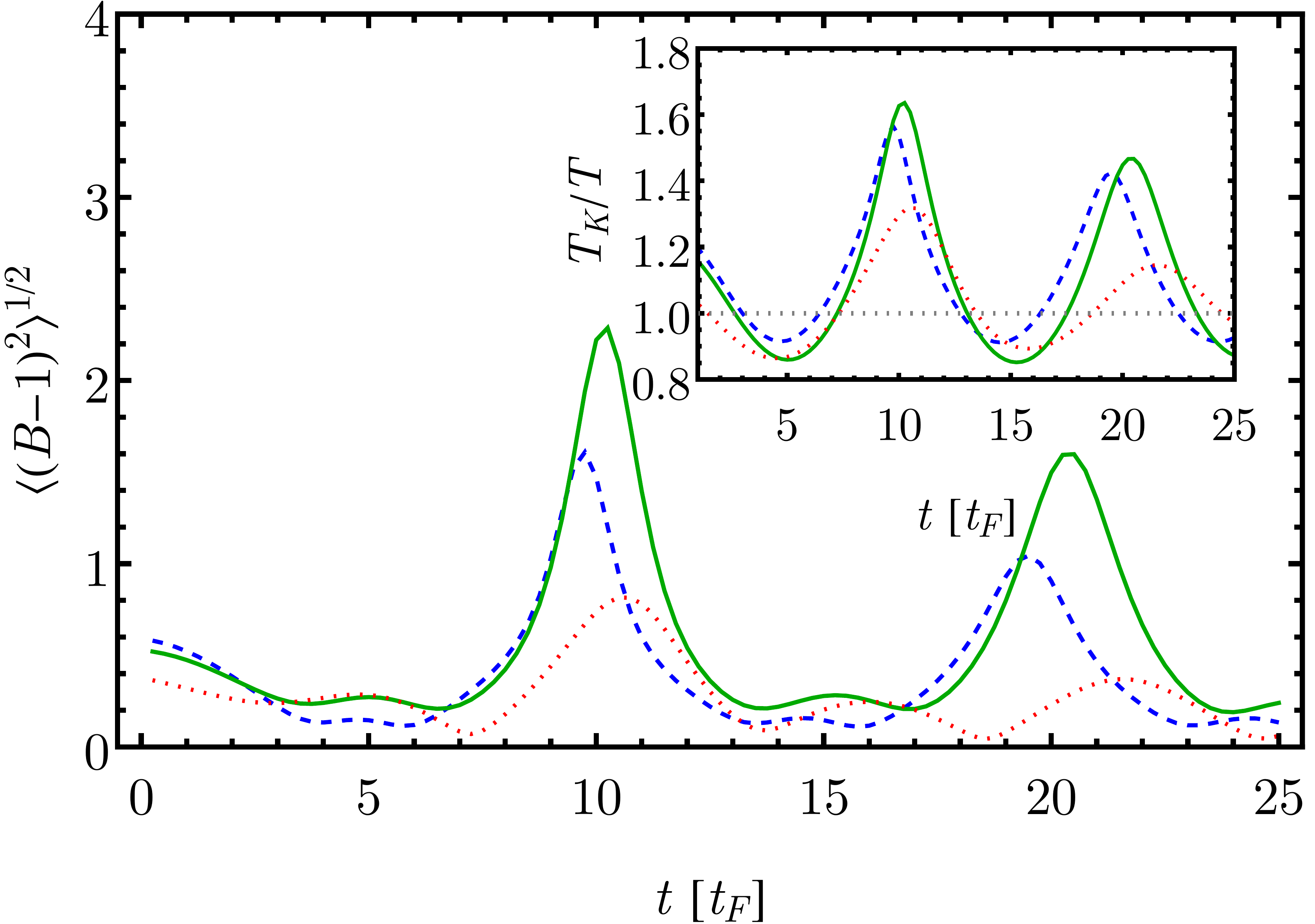}
\caption{Thermalization and violation of the detailed balance for attractive polarons. The main panel shows the degree of the violation of the detailed balance as quantified by $\surd\langle [B(\bk,\bk')-1]^2\rangle$. Inset shows the inferred temperature $T_K$ (in the units of the bath temperature $T$) from the Kullback-Leibler divergence. Here, the parameters are $1/k_Fa =$ -0.25 (dashed blue), 0 (solid green), 0.25 (dotted red); $\Omega=0.7\ve_F$, $T=0.135\ve_F$, $\delta_\att(0)=0$.
}
\label{fig_th}
\end{figure}

\textit{Thermalization.}---
While we have found that the effect of the collisional integral $I_\alpha$ is negligible in the Rabi oscillation dynamics, it drives the system to a thermal state by redistributing the population. We focus again on the attractive polaron case, and consider the question how far from equilibrium the distribution $n_\att(\bk, t)$ is. To probe the violation of  detailed balance, we define
\begin{equation}
    B \equiv \frac{1 - n_\att(\bk)}{n_\att(\bk)}
            \frac{1 - n_1(\bk_1)}{n_1(\bk_1)}
            \frac{n_\att(\bk')}{1 - n_\att(\bk')}
            \frac{n_1(\bk_1')}{1 - n_1(\bk_1')},
\end{equation}
where we dropped the dependence on $t$.  The two pairs of collision momenta
$(\bk',\bk_1') \rightleftarrows (\bk,\bk_1)$ satisfy energy and momentum conservation laws. The values $B>1$ ($B<1$) indicate that, in the kinetic equation for
$n_\att(\bk)$, the rate of \textit{in} processes is larger (smaller) than the \textit{out} processes in the collision integral in the chosen momentum sector.
In equilibrium, the \textit{in} and \textit{out} processes are in detailed balance and $B\equiv 1$ for all momenta. In our case, the bath
is held at equilibrium and $B$ becomes a function of $\bk$ and $\bk'$ only.

As an estimator of the lack of detailed balance we use $\surd\langle [B(\bk,\bk')-1]^2\rangle$, where the average is taken over the distribution $\propto n_\att(\bk,t)n_\att(\bk',t)$, whose dynamics is reported in 
Fig.~\ref{fig_th}. Our estimator shows a decaying oscillatory behaviour in time with pronounced peaks -- when the density of polarons is small -- where the detailed balance is significantly violated, and, the system is far from equilibrium (as can be seen also by looking at the momentum distribution itself -- see SM).

In order to investigate the dynamics of thermalization, we introduce the concept of the inferred temperature $T_K$, which we define based on the Kullback-Leibler divergence $D(P|Q)\equiv - \sum_\bk P_\bk \ln(Q_\bk/P_\bk)$, where $P$ and $Q$ are two probability distributions. This non-negative object, widely used in the theory of non-equilibrium processes~\cite{parrondo2009entropy, van2007dissipation}, which is nullified only if $P=Q$, quantifies the loss of information when the normalized distribution $Q_\bk$ is used for the approximation of the true distribution $P_\bk$; an optimization of the information loss is called as a moment projection~\cite{murphy2012machine}.
To search for the best approximation of the polarons in terms of equilibrium states, we take $P_\bk \propto n_\att(t,\bk)$ and $Q_\bk \propto n_\att^{\mathrm{eq}}(T^*,\bk)$, which is a thermal state at a guessed temperature $T^*$. The optimal estimate of the temperature is obtained by the minimization of the Kullback-Leibler divergence over $T^*$, assuming that the densities of the distribution are the same. This optimal temperature, which we denote with $T_K$, is the information projection to a set of thermal states, and reads
\begin{equation}
    T_K \equiv \underset{T^*}{\mathrm{arg\,min}}\ D\big(\tilde n_\att(t)\, \big|\, \tilde n_\att^\mathrm{eq}(T^*) \big),
\end{equation}
where the tilde denotes normalization with respect to momentum $\bk$. The chemical potential of equilibrium polarons $n_\att^\mathrm{eq}(T^*,\bk)$ is adjusted so that its density coincides with the one given by $n_\att(t,\bk)$.

In the inset of Fig.~\ref{fig_th}, we present the inferred $T_K$ relative to the temperature $T$ of the bath. We find that $T_K$ is a meaningful characteristic of the distribution, as it yields a continuous function with the correct order of magnitude.  The temperature $T_K$ shows oscillations on the experimental time scale, with the peak value on the order of 1.3$T$---1.6$T$  at times when the density of polarons is the lowest, i.e., at the maxima of the Rabi oscillations, cf. Fig.~\ref{fig_att}. In the long time limit, the temperature $T_K$ approaches the bath temperature. In the examples, the fastest thermalization is observed for the case $1/(k_F a) = 0.25$. 

Finally, close to thermal equilibrium, the violation of the detailed balance is small. We found that it is bounded by the Kullback-Leibler divergence, since for $n_\att(t) \approx n_\att^\mathrm{eq}$, we have 
\begin{equation}
    {\langle [B(\bk,\bk')-1]^2\rangle} \leqslant \zeta D(\tilde n_\att(t)|\tilde n_\att^{\mathrm{eq}}(T)),
\end{equation}
where $\zeta > 0$ is a function of thermodynamic variables at $T$. 
Thus, in the approach to equilibrium, the Kullback-Leibler divergence $D$ not only implies the inference of the correct temperature, i.e., $T_K \to T$, but also it sets a bound on the degree of the detailed balance violation. Our results show that on the timescale of the experiment, the system is far from the thermal state and a non-equilibrium description is required. More technical details on the thermalisation process is discussed in SM.

\textit{Conclusions and perspectives.}--- In this work, we provided  quantum kinetic equations for the single-particle density matrix of Fermi-polarons coherently driven between two internal levels. In the impurity limit, and as long as the polarons are well defined quasi-particles, our kinetic equations take the intuitive form shown in Eq.~(\ref{eq:keq}) -- a many-body version of a dressed two level system.
{An important feature of our approach is that it manifestly separates decoherence from momentum relaxation mechanisms, the latter entering the equations as a collision integral which redistributes particles between different momenta.
We compared our results with available experimental data for the ground state (attractive) polaron, finding good agreement without any fitting parameters. The theory is also applicable to the meta-stable (repulsive) polaron branch far from unitary limit.
Our approach provides a general tool to study out-of-equilibrium problems related to the fundamental concept of quasi-particles in many-body quantum systems, such as impurity thermalization, generation of quasi-particles in presence of strong Rabi coupling and repulsive-attractive polaron coherence, to mention a few.}

Finally, very recently a new experiment on Rabi dynamics was reported~\cite{nir2023strongly} on which our theory could be tested. In the same experiment a sharp transition between weak and large Rabi drive strength have been observed. Our approach can be applied to this situation provided we use the dressed Rabi particles as initial ingredients. Such route has been developed in \cite{levy1989spin} for Fermi liquid theory in large transverse magnetic fields.

\begin{acknowledgments}
\textit{Acknowledgments.}--- {We thank Matteo Zaccanti and Francesco Scazza for providing us with the experimental data.} Discussion with J. Levinsen are also acknowledged. Financial support from the Italian MIUR under
the PRIN2017 project CEnTraL (Protocol Number 20172H2SC4), from the Provincia Autonoma di Trento and from Q@TN, the joint lab between University of Trento, FBK-
Fondazione Bruno Kessler, INFN- National Institute for Nuclear Physics and CNR- National Research Council is acknowledged. 
This research is part of the project No. 2021/43/P/ST2/02911 co-funded by the National Science Centre and the European Union’s Horizon 2020 research and innovation programme under the Marie Skłodowska-Curie grant agreement no. 945339.
For the purpose of Open Access, the author has applied a CC-BY public copyright licence to any Author Accepted Manuscript (AAM) version arising from this submission.
\\

\textit{Data availability:} The data presented in this article is available from~\cite{data}.
\end{acknowledgments}

\end{document}


\author{Tomasz Wasak}\email{twasak@umk.pl}\affiliation{\NCU}\affiliation{\PKS}  
\author{Matteo Sighinolfi}\affiliation{\BEC}
\author{Johannes Lang}\affiliation{\PKS}\affiliation{\UK}
\author{Francesco Piazza}\email{piazza@pks.mpg.de}\affiliation{\AUG}\affiliation{\PKS}  
\author{Alessio Recati}\email{alessio.recati@ino.cnr.it}\affiliation{\BEC}\affiliation{\TIFPA}

\title{Supplemental Material: Decoherence and momentum relaxation in Fermi-polaron Rabi dynamics: a kinetic equation approach}
\maketitle

In this Supplemental Material (SM), we present derivations of the kinetic equations (3)
of the main text. To this aim, we write the theory in the Keldysh formalism and then we approximately solve the Dyson equations to obtain the desired kinetic equations. We also include a derivation of the collision integral $I_\alpha$ and of the polaron decoherence rate $\Gamma_\alpha^\mathrm{dec}$. Finally, we also discuss the repulsive polaron case. 
For more details on the Keldysh formalism and on its usage, we refer the reader to \cite{kamenev2011field,wasak2021quantum,wasak2021fermi} and references therein. A more detailed derivation of the kinetic equations can be found in Ref.~\cite{Sighinolfi2022}.

\section{Keldysh formulation of the theory}

In our work, we use the Keldysh formalism as presented in~\cite{kamenev2011field} to describe our theory. In this way, we can compute the expectation value of any operator using the path integral formalism. 

The total Hamiltonian $\hat H$ of the system is given by Eq.~(1)
of the main text. The Keldysh partition function ${\cal Z}$ of $\hat H$ is
\begin{equation}
    {\cal Z} = \int \! {\cal D}\psi_1{\cal D}\psi_2{\cal D}\psi_3~e^{i S[\psi_{1+},\psi_{1-},\bar{\psi}_{1+},\bar{\psi}_{1-},\psi_{2+},\psi_{2-},\bar{\psi}_{2+},\bar{\psi}_{2-},\psi_{3+},\psi_{3-},\bar{\psi}_{3+},\bar{\psi}_{3-}]},
\end{equation}
where the integration measures are in the form ${\cal D}\psi_i=\prod_{\sigma = \pm}{\cal D}\bar\psi_{i\sigma}{\cal D}\psi_{i\sigma}$.
The fields $\psi_{i\sigma}$ are Grassmann anticommuting fields for the fermions in level $i$ and the $+(-)$ index labels fields on the forward (backward) branch of the Keldysh time contour~\cite{kamenev2011field}. The total action can be decomposed, analogously to $\hat H$, as $S=\sum_i S_i + S_\Omega + S_{\rm int}$, i.e., free particle actions $S_i$, the Rabi action $S_\Omega$ and the interaction part $S_{\rm int}$.  
After the Keldysh rotation of the fermionic fields defined as~\cite{kamenev2011field,larkin1975nonlinear}
\begin{align}
    \psi_{i1} &= \frac{1}{\sqrt{2}}(\psi_{i+}+\psi_{i-}), \quad \psi_{i2}=\frac{1}{\sqrt{2}}(\psi_{i+}-\psi_{i-}) \label{keld rot psi}\\
    \bar\psi_{i1} &= \frac{1}{\sqrt{2}}(\bar\psi_{i+}- \bar\psi_{i-}), \quad \bar\psi_{i2}=\frac{1}{\sqrt{2}}(\bar\psi_{i+}+ \bar\psi_{i-}),
\end{align}
the free action component is (summation over repeated indices $a,b$ is understood from now on; each index takes components $1$ or $2$ in the Keldysh space) 
\begin{equation}
    S_i = 
    \iint \! dx dx'~\bar\psi_{i a}(x) \hat{G}^{-1}_{0i,ab} (x,x') \psi_{i b}(x'),
\end{equation}
with $dx = d{\bf r}dt$ and $\hat{G}^{-1}_{0i}$ is the $2\times 2$ bare inverse matrix of Green functions (GF); summation over repeated indices is implied. 

\begin{figure}[t]
\centering
    \includegraphics[width=0.48\textwidth,keepaspectratio=true]{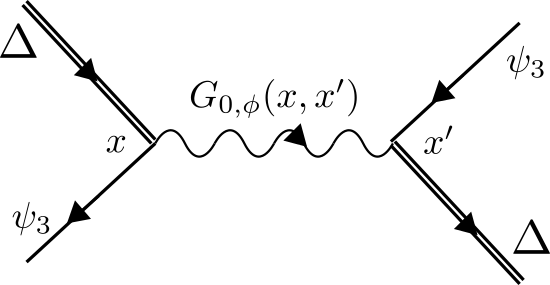}
\caption{\label{Fig: vertex}Vertex of $S_{\Delta 3}$ in position space as represented in Eq.~\eqref{effective action}. Each line carries a Keldysh index, with $\alpha, \, \beta = cl, \, q$ and $c, \, d = 1 \, 2$. }
\end{figure}

For our fermionic system, the causal structure of this matrix is given by, 
\begin{equation}
    \hat{G}^{-1}_{0i}=\begin{pmatrix} 
    G^{-1,R}_{0i} & G^{-1,K}_{0i} \\
    0 & G^{-1,A}_{0i}
    \end{pmatrix}.
\end{equation}
In the last expression, the dependence on $(x,x')$ is understood and R, A, K label respectively the retarded, advanced and Keldysh components of the GF.
With the same convention, the Rabi action is 
\begin{equation}
S_\Omega = \frac{\Omega}{2} \int \! dx ~ \big[ \bar\psi_{3a}(x)\psi_{2a}(x)+\bar\psi_{2a}(x)\psi_{3a}(x) \big].
\end{equation}
Note that, differently from the main text, here we reabsorb the detuning $\Delta$ of level 2 inside the free particle action $S_2$. The non-Keldysh-rotated interaction action is
\begin{equation}
    S_{\rm int} = -U \sum_{\sigma = \pm} \int \! dx~ \bar\psi_{3\sigma}(x) \bar\psi_{1\sigma}(x) \psi_{1\sigma}(x) \psi_{3\sigma}(x).
\end{equation}
$S_{\rm int}$ is quadratic in the bath fields $\psi_1$ and in order to obtain an action that is linear in $\psi_1$ we introduce via a Hubbard-Stratonovich transformation (HST) an auxiliary molecular field $\Delta_\sigma(x)=U\psi_{1\sigma}(x)\psi_{3\sigma}(x)$ with bosonic statistics. The HST is explicitly written as
\begin{equation}
    e^{iS_{\rm int}}=\int D[\bar{\Delta},\Delta]e^{iS_{\Delta}+iS_{\Delta \psi_1 \psi_3}}, 
\end{equation}
with the actions $S_\Delta$ and $S_{\Delta \psi_1 \psi_3}$ defined on the Keldysh contour ${\cal C}$ 
\begin{align}
&S_{\Delta}= \int_{\cal C} \! dx ~  \bar{\Delta}(x) \frac{1}{U} \Delta(x) \\
&S_{\Delta \psi_1 \psi_3}=-\int_{\cal C} \! dx ~ [\bar{\psi}_3\bar{\psi}_1\Delta + \bar{\Delta}\psi_1 \psi_3].
\end{align}
Now $S$ is linear in the bath field $\psi_1$ as desired, therefore they can be integrated out using standard Gaussian integrals. The final effective action, after the proper Keldysh rotation, reads
\begin{equation}\label{effective action}
S_{\Delta 3}=-\iint \! dx ~ dx' ~ \bar{\zeta}_a(x)\hat G_{01,ab}(x,x')\zeta_b(x'), 
\end{equation}
where the fields $\zeta$ are defined as

\begin{equation}
\zeta_1=\frac{\bar{\psi}_{3,1}\Delta_2+\bar{\psi}_{3,2}\Delta_1}{\sqrt{2}} \quad \zeta_{2}=\frac{\bar{\psi}_{3,1}\Delta_1+\bar{\psi}_{3,2}\Delta_2}{\sqrt{2}} \quad
\bar{\zeta}_1=\frac{\bar{\Delta}_1\psi_{3,2}+\bar{\Delta}_2\psi_{3,1}}{\sqrt{2}} \quad \bar{\zeta}_2=\frac{\bar{\Delta}_1\psi_{3,1}+\bar{\Delta}_2\psi_{3,2}}{\sqrt{2}},
\end{equation}
with the Keldysh rotation of $\Delta$ defined as in Eq.\eqref{keld rot psi}, while $\bar\Delta$ is transformed by
\begin{equation}
    \bar\Delta_1=\frac{1}{\sqrt{2}}(\bar\Delta_{+} + \bar\Delta_{-}) \quad \bar\Delta_2=\frac{1}{\sqrt{2}}(\bar\Delta_{+} - \bar\Delta_{-}).
\end{equation}
The different rotations for $\bar\Psi$ and $\bar\Delta$ reflect different statistics of the fields.  
In Eq.~\eqref{effective action}, we now have an interaction between $\psi_3$ and $\Delta$ mediated by the fermions in state 1. This describes a scattering between an impurity in level 3 and a molecule, as shown in Fig.\ref{Fig: vertex}.

\section{Derivation of Dyson equations for Keldysh Green functions}
Dressed GFs for molecules and impurities can be derived from Dyson equations in the Keldysh formalism, with the key difference that the $2\times2$ structure of the GF matrix in this formalism gives 3 different equations to solve for the R, A, K components. The retarded GF is needed to describe the excitation spectrum, while the Keldysh GF is related to the distribution function $F$ via $G^K = G^R \circ F - F \circ G^A$, where the symbol $\circ$ stands for the convolution operator~\cite{kamenev2011field}.  

\begin{figure}[t]
  \centering
  \includegraphics[width=0.48\textwidth]{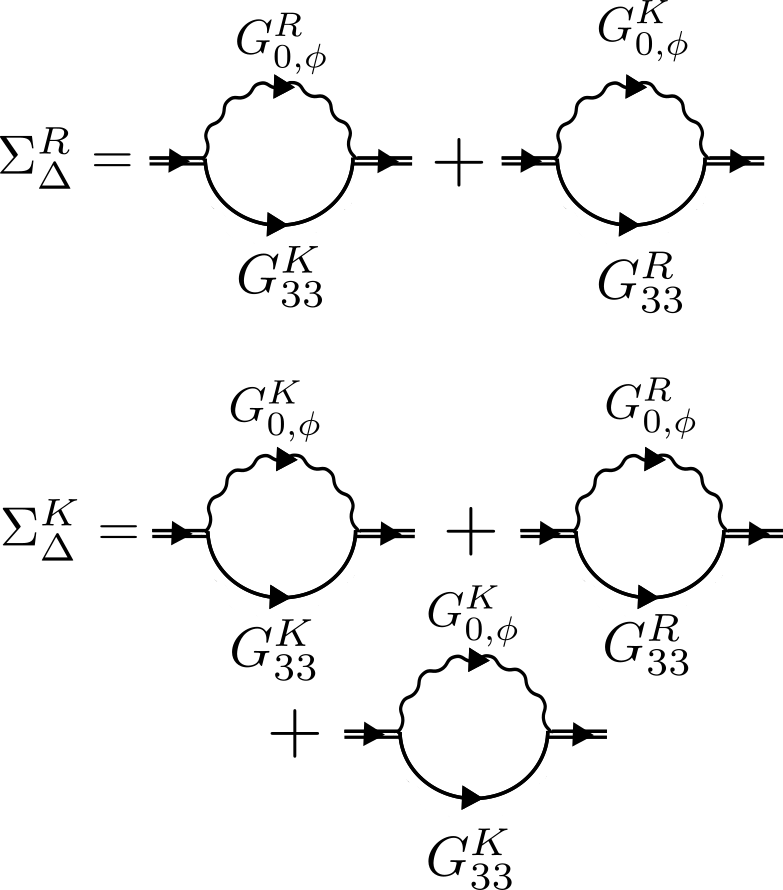}
  \includegraphics[width=0.48\textwidth]{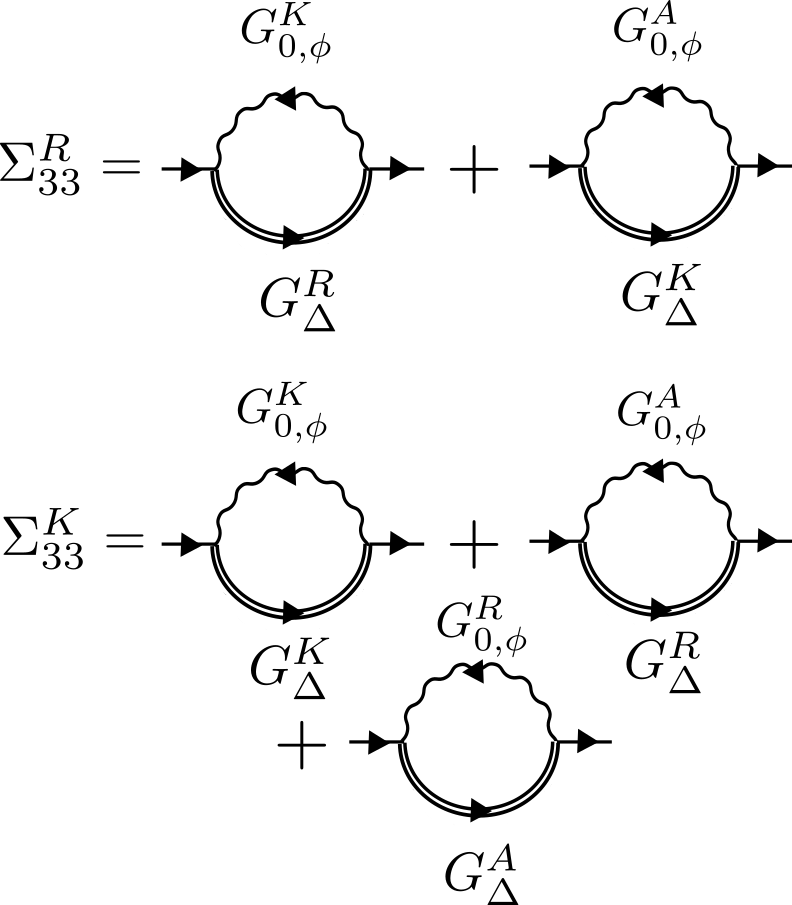}
\caption{\label{Fig:self_energies} Schematic representation of the molecular and impurity self energies, $\Sigma_\Delta$ and $\Sigma_{33}$, defined in Eqs.~\eqref{sig del r}-\eqref{skwt}. Wavy lines denote $G_{01}$, full lines $G_{33}$ and double lines $G_\Delta$, while the labels $R$, $A$, $K$ denote the components of the propagators inside the loop.}
\end{figure}

The Dyson equations for molecules and for atoms in levels 2 and 3 read
\begin{align}
    (\hat G_{0\Delta}^{-1} - \hat\Sigma_\Delta) \cdot \hat G_\Delta &= \mathbb{1} \label{Dyson Delta} \\
    (\hat G_{0, i j}^{-1} - \hat\Sigma_{i j}) \cdot \hat G_{j k} &= \mathbb{1}\delta_{ik}, \label{Dyson imp} \quad  \\ 
\end{align}
where $i,j,k=2,3$ and the dot product denotes the matrix product combined with a convolution. In the last equation, we adopted for the bare and dressed GF of atoms the same spin notation as in Ref.~\cite{kamenev2011field}. The bare $G_0^{-1 \, K}$ are regularization factors and can therefore be neglected for both atoms and molecules. The bare retarded and advanced Green's functions are

\begin{align}
G_{0 \Delta}^{-1 \, R(A)}(x_1,x_2) &= \delta(x_1-x_2) U \label{GF Delta bare} \\ 
G_{0, j k}^{-1 \, R(A)}(x_1,x_2) &= \delta(x_1-x_2)\left[\left(i\partial_{t_2} + \frac{1}{2m}\nabla^2_{\mathbf{r}_2}\right)\delta_{j k}-\mathbf{H}\cdot \hat{s}_{j k}+i0^{+}\delta_{jk}  \right] \, , \label{GF fermion bare} 
\end{align}
where $\hat{s}$ is the vector of Pauli matrices and $\mathbf{H}=(\Omega/2,0,\Delta/2)$, consistent with the definition of $H_\Omega$. Note that $\Omega$ is in general time dependent. The time-dependence has to be chosen in order to match the experimental conditions described in Ref.~\cite{scazza2017repulsive} and the related Supplemental Material. This means that first the detuning $\Delta$ is tuned in resonance with the desired polaron branch and then the Rabi frequency is switched on at the initial time of the dynamics.
To solve the Dyson equation we compute the 1-loop self-energies $\hat \Sigma_{j k}$ and $\hat\Sigma_\Delta$. These are obtained by contracting the vertex of $S_{\Delta 3}$ with a Green's function. Following the summation over internal Keldysh indices one finds
\begin{align}
    \Sigma_{\Delta}^R(p) &= \frac{i}{2V} \sum_q \left( G_{01}^R(p-q)G_{0,33}^K(q)+G_{01}^K(p-q)G_{0,33}^R(q)  \right) \label{sig del r} \\
\Sigma_{\Delta}^K(p) &= \frac{i}{2V} \sum_q \left\lbrace G_{01}^K(p-q)G_{0,33}^K(q)+ [G_{01}^R(p-q)-G_{01}^A(p-q)] [G_{0,33}^R(q)-G_{0,33}^A(q)] \right\rbrace \label{sig del k} \\
\Sigma_{33}^R(x,p)&=-\frac{i}{2V} \sum_q \left( G_{01}^K(q)G_{\Delta}^R(x,p+q) + G_{01}^A(q)G_{\Delta}^K(x,p+q) \right) \label{srwt} \\
\Sigma_{33}^K(x,p)&=-\frac{i}{2V} \sum_q \left( G_{01}^K(q)G_{\Delta}^K(x,p+q) + G_{01}^A(q)G_{\Delta}^R(x,p+q) + G_{01}^R(q)G_{\Delta}^A(x,p+q) \right) \label{skwt}, 
\end{align}
with $x=(x_1+x_2)/2$ and $p=(\omega,{\bf p})$ is the relative momentum, i.e., the conjugate coordinate of $x_1-x_2$ in the sense of the Fourier transform. In our approximation scheme, we compute $\Sigma_\Delta$ using the bare GF, Eqs. (\ref{GF Delta bare})-(\ref{GF fermion bare}): since the bare GF are time and space-translation invariant, the molecular self energy depends only on the relative momentum. We then use the molecular self energies obtained in this way to derive --- as explained below --- dressed molecular propagators $G_\Delta$ that we use to compute $\Sigma_{33}$, Eqs. (\ref{srwt})-(\ref{skwt}), that also depends on $x$. In the impurity limit, i.e., when the density of the impurities is vanishingly small compared to the density of the bath, this scheme is equivalent to the non-self-consistent $T$-matrix approximation for $\Sigma_{33}$. A representation of the self-energies is shown in Fig.~\ref{Fig:self_energies}.

\section{Retarded Green functions and polaron energy}

The first quantity we need to know, because it is needed for $\hat\Sigma_{33}$ in our approximation scheme, is $\gd$. From the Dyson equation for the molecules, Eq.~\eqref{Dyson Delta}, we obtain

\begin{equation}
    G_\Delta^R(p) = \frac{1}{U^{-1} - \Sigma_\Delta^R(p)}
\end{equation}
The explicit form of $G_\Delta^R$ is obtained through the relations $G_i^K(p)=F_i(p)(G_i^R(p)-G_i^A(p))$. For the bath at equilibrium, $F_1(p) = 1-2n_1^{\rm eq}({\bf k})$, with $n_1^{\rm eq}$ the Fermi-Dirac distribution at temperature $T$, while in the impurity limit $N_3 \ll N_1$, $F_3 \simeq 1$. The form of $G_\Delta$ obtained in this way is
\begin{equation}
\gd (\omega, \bk)=\left(\frac{m}{4\pi a_{1 3}} -\frac{1}{V}\sum_{\bq} \frac{1-n_1^{\rm eq}(\bk - \bq)}{\omega-\varepsilon_3(\bq)-\varepsilon_1(\bk - \bq)+i0^+ }+\frac{m}{\bq^2} \right)^{-1}, \label{retard mol prop}
\end{equation}
where the contact interaction strength $U$ is related to the $s$-wave scattering length $a_{13}$ by the Lippmann-Schwinger equation, so that $U^{-1}=m/4\pi a_{13}+\sum_{\bq}m/\bq^2$~\cite{chevy2006universal}. 

\begin{table}[t]
    \parbox{.45\linewidth}{
    \begin{tabular}{c|c|c|c|c}
        $1/k_F a$ & $E_0$ $[\varepsilon_F]$ & $m_*/m$  & $Z_a(0)$ & $\partial_\omega \im \sig(0) $ \\
        \hline
        0 & -0.625 & 1.16 & 0.775 & -69.0 $\times 10^{-4}$ \\
        0.25 & -0.858 & 1.29 & 0.673 & -33.5 $\times 10^{-4}$ \\
    \end{tabular}
    \caption{Attractive polaron parameters at $T=0.135 T_F$. }
    \label{tab:Attractive polaron}}
    \parbox{.45\linewidth}{
    \begin{tabular}{c|c|c|c|c}
        $1/k_F a$ & $E_0$ $[\varepsilon_F]$ & $m_*/m$ & $Z_r(0)$ & $\partial_\omega \im \sig(0)$ \\
        \hline
        1.27 & 0.423 & 1.13 & 0.823 & -2.7 $\times 10^{-4}$ \\
        2.63 & 0.188 & 1.07 & 0.963 & -0.3 $\times 10^{-4}$
    \end{tabular}
    \caption{Repulsive polaron parameters at $T=0.135 T_F$}
    \label{tab:Repulsive polaron}}
    \parbox{.45\linewidth}{
     \begin{tabular}{c|c}
        $1/k_Fa$ & $1-\sum_\alpha Z_\alpha$  \\
        \hline
        0.0 & 0.225 \\
        0.25 & 0.140 \\
        1.27 & 0.037 \\
        2.63 & 0.018
    \end{tabular}
    \caption{Sum of quasiparticle residues at $T=0.135$.}
    \label{tab:quasiparticles residues}
    }
\end{table}

The same approach applied to $G_{33}^R$ gives 
\begin{equation}
G_{33}^{R}(x,p)= \left( \omega-\varepsilon_3(\bk)-\Sigma_{33}^{R}(x,p)-\frac{\Omega^2/4}{\omega-\varepsilon_2(\bk)+\Delta + i0}\right)^{-1}, \label{dressed g33 rit}
\end{equation}
where the last term describes the Rabi coupling to state 2. The polaron energy is the pole of the propagator, and the Rabi coupling splits each branch of the polaron spectrum in two sub-branches. Indeed, the $\alpha$-polaron dispersion relation is  

\begin{equation}\label{polaron dispersion relation complete}
\varepsilon_{\alpha \, 1,2}(\bk)=\varepsilon_3(\bk)+\frac{\re \Sigma_{33}^R (\varepsilon_{\alpha \, 1,2}(\bk),\bk)-\Delta}{2}\pm \frac{1}{2}\sqrt{(\re \Sigma_{33}^R (\varepsilon_{\alpha \, 1,2}(\bk),\bk)+\Delta)^2+\Omega^2 } .
\end{equation} 
However, we will not consider the dressing effect of the Rabi coupling on the branches in our treatment, leaving it for future development.
Therefore, the polaron energy dispersion relation has the usual form
\begin{equation}\label{polaron dispersion relation approx}
    \varepsilon_\alpha(\bk) = \varepsilon_3(\bk) + \re \Sigma_{33}^R(\varepsilon_\alpha(\bk),\bk).
\end{equation}
 
Finally, a comment on $\re \sig$ is also useful: 
we derive it in the impurity limit, where the non-self consistent $T$-matrix approximation is recovered as known for equilibrium~\cite{combescot2007normal}. Within this approximation polaron dispersion relation, quasiparticle residue $Z_\alpha (\bk = 0)$ and effective mass $m_*$ are derived. Results obtained in the non-self consistent $T$-matrix approximation are known to be in good agreement with experiments and quantum Monte Carlo calculations, see e.g. Ref.~\cite{massignan2014polarons} and references therein. The results used in the numerical simulation are listed in Tabs.~\ref{tab:Attractive polaron} and ~\ref{tab:Repulsive polaron}. As we will show below, we have to go beyond the impurity limit to describe the dynamics of populations.

\section{Kinetic equations, collision integrals and decoherence rates}
The last component of the Dyson equation is the Keldysh one, and it is necessary to derive kinetic equations. We start our derivation from the molecular Dyson equation, where the Keldysh component is
\begin{equation}\label{boltzmann delta 0}
G_\Delta^{-1 \, R} \circ G_\Delta^K \circ G_\Delta^{-1 \, A} = \Sigma_\Delta^K - G_{0,\Delta}^{-1 \, K}.
\end{equation}

We then {use} the relation $G_{\Delta}^K=G_{\Delta}^R \circ F_{\Delta} - F_{\Delta} \circ G_{\Delta}^A$ for the distribution function $F_\Delta$ {and we insert it} into Eq.~\eqref{boltzmann delta 0} to obtain an equation for the distribution function of molecules that reads  
\begin{equation}\label{boltzmann delta 1}
 F_{\Delta}\circ G_{\Delta}^{-1 \, A} - G_{\Delta}^{-1 \, R} \circ F_{\Delta} = \Sigma_{\Delta}^K - G_{0,\Delta}^{-1 \, K} .
\end{equation}
We then apply the Wigner transformation~\cite{kamenev2011field,sieberer2016keldysh} to simplify the equation for $F_\Delta$. We consider only the zeroth order of the expansion in gradients, treating the dynamics of the molecules as subordinated to the dynamics of the polarons. In this approximation, Eq.~\eqref{boltzmann delta 1} reads
\begin{equation}
2i F_{\Delta} \im \Sigma_{\Delta}^R = \Sigma_{\Delta}^K,
\end{equation}
and the following equations hold:
\begin{align}
&F_{\Delta}(G_{\Delta}^R-G_{\Delta}^A)=|G_{\Delta}^R|^2\Sigma_{\Delta}^K \label{inst sol 1} \\
&F_{\Delta}{\cal A}_\Delta = |G_{\Delta}^R|^2 i \Sigma_{\Delta}^K \label{inst sol 2}\\
&{\cal A}_\Delta=-2|G_{\Delta}^R|^2 \im \Sigma_{\Delta}^R, \label{inst sol 3} 
\end{align}
where ${\cal A}_\Delta = -2 \im \gd$.

Now we consider the impurities. From Eq.~\eqref{Dyson imp} we have 4 kinetic equations for $F_{22}$, $F_{33}$, $F_{23}$ and $F_{32}$ but only 3 of them are relevant, because $F_{32}=F_{23}^*$. It is convenient to define $\tilde G_{0,jk}^{-1 \, R(A)}$ as
\begin{equation}\label{G0 generic}
\tilde G_{0,jk}^{-1 \, R(A)}(x_1,x_2)=\delta(x_1-x_2)\delta_{jk} \left(i\partial_{t_2} + \frac{1}{2m}\nabla^2_{\mathbf{r}_2}  \right),
\end{equation}
so that $G_{0,jk}^{-1 \, R(A)}= \tilde G_{0,jk}^{-1 \, R(A)} - \mathbf{H}\cdot \hat{s}_{j k}$

 From the matrix structure of the Dyson equation for the impurities we have that
\begin{align}
    &F_{22} \circ \tilde G_{0,22}^{-1 \, A} - \tilde G_{0,22}^{-1 \, R} \circ F_{22} + \frac{\Omega}{2} \circ F_{23}^* - F_{23} \circ \frac{\Omega}{2} = 0, \\
    & F_{33} \circ \tilde G_{0,33}^{-1 \, A} - \tilde G_{0,33}^{-1 \, R} \circ F_{33}  + \frac{\Omega}{2} \circ F_{23} - F_{23}^* \circ \frac{\Omega}{2} + \Sigma_{33}^R \circ F_{33} - F_{33} \circ \Sigma_{33}^A = \Sigma_{33}^K, \\
    & F_{23} \circ \tilde G_{0,33}^{-1 \, A} - \tilde G_{0,22}^{-1 \, R} \circ F_{23} + \frac{\Omega}{2} \circ F_{33} - F_{22} \circ \frac{\Omega}{2} + \Delta F_{23}  = F_{23} \circ \Sigma_{33}^A. %
\end{align}
We now perform a Wigner transformation at linear order in gradients on the previous equations. Note that the free fermionic energies $\ve_2$ and $\ve_3$ are the same and this leads to (the dependence $F_{ij}(x,k)$ is not shown)

\begin{align}
    &\left\lbrace \omega - \varepsilon_2(\bk), \, F_{22} \right\rbrace - i\frac{\Omega}{2}(F_{23}-F_{23}^*)-\frac{1}{2}\left\lbrace \frac{\Omega}{2},F_{23}+F_{23}^* \right\rbrace=0 \label{wigner 22} \\ 
    &\ \left\lbrace \omega - \varepsilon_3(\bk) - \re\sig (x,k), \, F_{33} \right\rbrace + i\frac{\Omega}{2}(F_{23}-F_{23}^*) -\frac{1}{2}\left\lbrace \frac{\Omega}{2},F_{23}+F_{23}^* \right\rbrace = i\Sigma_{33}^K(x,k) +2\im \sig (x,k)F_{33} \label{wigner 33} \\
    & \left\lbrace \omega - \varepsilon_3(\bk) - \frac{1}{2}\re\sig (x,k), \, F_{23} \right\rbrace -i ( \re \sig (x,k)-\Delta)F_{23} -i \frac{\Omega}{2}(F_{22}-F_{33}) -\frac{1}{2}\left\lbrace \frac{\Omega}{2},F_{22}+F_{33} \right\rbrace + \nonumber \\
    &\hspace{5.5 cm} +\frac{i}{2}\lbrace \im \Sigma_{33}^R (x,k) , F_{23} \rbrace   =\im \sig (x,k)F_{23},  
    \label{wigner 23}
\end{align}
where the Poisson brackets are defined as $\lbrace A,B \rbrace=  \partial_x A \partial_k B - \partial_k A \partial_x B $ and $\partial_x A \partial_k B = \nabla_{\mathbf{r}}A \nabla_{\bk}B - \partial_t A \partial_{\omega}B$.

Note that the right-hand side of the Eq.~\eqref{wigner 33} defines the collision integral $\tilde I_{\rm coll} = i\Sigma_{33}^K + 2\im\sig F_{33}$, while in the right-hand side of Eq.~\eqref{wigner 23} we only have the imaginary part of the retarded self-energy, so this term is a decay rate. We also point out that we are not in the impurity limit: indeed a small but finite concentration of impurities is necessary to compare with experimental results. Note that in Eqs.~(\ref{wigner 22})-(\ref{wigner 23}) the explicit time dependence of the Rabi frequency is $\Omega(t)=\theta(t)\Omega$, but it will be not considered form now on~\footnote{Even if this time dependence is included, it does not change 
the final kinetic equations because it will be removed by the procedure adopted.}.

The equations above describe the distribution functions $\fij{ij}$ for any $\omega$ 
and we do not have the equations for populations yet. In systems with a Rabi coupling,  
the connection between $\fij{ij}$ and the populations of the levels is not trivial 
to find, as opposed to the common situation in 
literature~\cite{kamenev2011field,ruckenstein89,sieberer2016keldysh}.

A way to find this connection is to work with the Kadanoff-Baym equation~\cite{kadanoff2018quantum} 
for the Keldysh component of the GF. This equation is, in matrix form 
\begin{align}
    &D_0 G^{\rm K} - G^{\rm K}D_0 + \frac{i}{2}(\Gamma G^{\rm K} + G^{\rm K} \Gamma) 
    +\frac{i}{2}\left[\left\lbrace D_0,G^{\rm K} \right\rbrace - \left\lbrace G^{\rm K}, D_0 \right\rbrace \right] 
    -\frac{1}{4}\left[\left\lbrace \Gamma,G^{\rm K} \right\rbrace + \left\lbrace G^{\rm K}, \Gamma \right\rbrace \right] = \nonumber \\
    &\hspace{3cm}=\Sigma^{\rm K}G^{\rm A}-G^{\rm R}\Sigma^{\rm K}+\frac{i}{2}\left[\left\lbrace \Sigma^{\rm K},G^{\rm A}-\left\lbrace G^{\rm R},\Sigma^{\rm K} \right\rbrace \right\rbrace \right],
\end{align}
where 
\begin{align}
    D_0 &= \begin{pmatrix}
        &\omega - \ve_{2,\bp} &-\Omega/2 \\
        &-\omega /2 &\omega-\ve_{3,\bp}-\re\sig+\Delta
    \end{pmatrix} \\ 
    \Gamma &= \begin{pmatrix}
        &0 &0 \\
        &0 &-2\im\sig
    \end{pmatrix} = \begin{pmatrix}
        &0 &0 \\
        &0 &\Gamma_{33}
    \end{pmatrix} .
\end{align}
When written in components, 
Kadanoff-Baym equation leads to equation that are similar to the 
ones for the distribution functions
\begin{align}
    &\partial_t G^{\rm K}_{22}+i\frac{\Omega}{2}(G^{\rm K}_{23}-G^{\rm K}_{32}) =0 \label{kb 22 interm} \\
    &(1-\dw \re \sig)\partial_t G^{\rm K}_{33} - i \frac{\Omega}{2}(G^{\rm K}_{23}-G^{\rm K}_{23}) = i\Sigma_{33}^{\rm K}(G_{33}^{\rm R}-G_{33}^{\rm A}) \label{kb 33 interm} \\
    &\left(1-\frac{1}{2}\dw \re \sig \right) \partial_t G^{\rm K}_{23} -i ( \re \sig -\Delta)G^{\rm K}_{23} +i \frac{\Omega}{2}(G^{\rm K}_{22}-G^{\rm K}_{33}) + \nonumber \\
    &\hspace{2.5 cm} +\frac{i}{2}\dw \im \sig \partial_t G^{\rm K}_{23} = -\frac{\Gamma_{33}}{2} G^{\rm K}_{23}+iG_{23}^{\rm R}\Sigma_{33}^K. \label{kb 23 interm}
\end{align}
At linear order in gradients of the WT, the matrix $\hat{G}^{\rm K}$ can be parametrized 
as 
\begin{equation}
    \hat{G}^{\rm K}=-\frac{i}{2}(\hat{\cal A}\hat{F}-\hat{F}\hat{\cal A}),
\end{equation}   
where $\hat{\cal A}$ is the matrix of spectral functions. In the limit of small 
$\Omega$, i.e. when the Rabi coupling is a probe, the energies of levels $\ket{2}$ 
and $\ket{3}$ are not modified and the components of $\hat{G}^{\rm K}$ are 
\begin{align}
    G_{22}^{\rm K}&=-i{\cal A}_{22}\fij{22} \\
    G_{23}^{\rm K}&=-\frac{i}{2}({\cal A}_{22}+{\cal A}_{33})\fij{23} \\
    G_{32}^{\rm K}&=-\frac{i}{2}({\cal A}_{22}+{\cal A}_{33})\fij{32} \\
    G_{33}^{\rm K}&=-i{\cal A}_{33}\fij{33},
\end{align} 
and in the same limit the spectral functions are 
\begin{align}
    {\cal A}_{22}&=2\pi\delta(\omega-\ve_2(\bp)) \\
    {\cal A}_{33}&=\sum_{\alpha= a,r}Z_\alpha 2\pi\delta(\omega-\ve_\alpha(\bp)), \label{spectral 33}
\end{align}
where $\ve_2$ is the bare energy of level $\ket{2}$ and $\ve_\alpha$ is given by the polaron energy not modified by the Rabi coupling, Eq.~\eqref{polaron dispersion relation approx}. With the above results for spectral functions and components of $\hat{G}^{\rm K}$, it is possible to perform and integration over $\omega$ and obtain the so-called on-shell equation for the distribution functions. The advantage of using the Kadanoff-Baym equation instead of the Dyson's equation is that for the former the on-shell projection is naturally defined, because all the distribution functions in the equations are multiplied by a spectral function. In the on-shell equations, also the terms with $\dw \im \sig$ is neglected because they are small, see Tables~\ref{tab:Attractive polaron} and \ref{tab:Repulsive polaron}. The final result is 
\begin{align}
    &\partial_t \fij{22}^{(2)} -i\frac{\Omega}{2}\left[\frac{1}{2}(\fij{23}^{(2)}-\fij{32}^{(2)})
    +\frac{1}{2}\sum_\alpha Z_\alpha(\fij{23}^{(\alpha)}-\fij{32}^{(\alpha)})\right] \\
    &\sum_\alpha \partial_t \fij{33}^{(\alpha)} +i\frac{\Omega}{2}\left[\frac{1}{2}(\fij{23}^{(2)}-\fij{32}^{(2)})
    +\frac{1}{2}\sum_\alpha Z_\alpha(\fij{23}^{(\alpha)}-\fij{32}^{(\alpha)})\right] = 
    \sum_\alpha (i\Sigma_{33}^{{\rm K}\, (\alpha)}+2\im\Sigma_{33}^{{\rm R}\, (\alpha)}\fij{33}^{(\alpha)}) \label{kb 33 on-shell F} \\
    &\left(1-\frac{1}{2}\dw \re \Sigma_{33}^{{\rm R}\, (2)}\right)\frac{1}{2}\partial_t \fij{23}^{(2)} +\sum_\alpha \tilde{Z}_\alpha^{-1}Z_\alpha\frac{1}{2}\partial_t \fij{23}^{(\alpha)} 
    -i\left[(\re\Sigma_{33}^{{\rm R}\, (2)} -\Delta)\frac{1}{2}\fij{23}^{(2)} +\sum_\alpha \times \right. \nonumber \\
    &\left. \times (\re\Sigma_{33}^{{\rm R}\, (\alpha)} -\Delta)\frac{1}{2}Z_\alpha\fij{23}^{(\alpha)} \right] 
    +i\frac{\Omega}{2}(\sum_\alpha Z_\alpha \fij{33}^{(\alpha)}-\fij{22}^{(2)}) = \frac{1}{2}\left(\im \Sigma_{33}^{(R) \, (2)}\fij{23}^{(2)}+\sum_\alpha Z_\alpha \im \Sigma_{33}^{(R) \, (\alpha)}\fij{23}^{(\alpha)}  \right), \nonumber \\
\end{align} 
where in the last equation the term $iG_{23}^{\rm R}\Sigma_{33}^{\rm K}$ vanishes on shell. The notation 
\begin{align}
    \fij{ij}^{(n)} &\equiv \fij{ij}(\omega = \ve_n) \\
    Z_\alpha &=  (1-\partial_\omega \re \sig |_{\omega = \ve_\alpha} )^{-1} \\
    \tilde{Z}_\alpha &= \left( 1-\frac{1}{2}\partial_\omega \re \sig |_{\omega = \ve_\alpha} \right)^{-1},
\end{align}
has been used.
Now, from the above equations for the on-shell distribution functions it is 
possible to obtain the ones for the populations. Indeed, in general one has that 
\begin{align}
    i\int_\omega G_{22}^{\rm K} &\equiv 1-2n_{22}^{(2)} = \fij{22}^{(2)} \\
    i\int_\omega G_{33}^{\rm K} &\equiv \sum_\alpha 1-2n_{33}^{(\alpha)} = \sum_\alpha Z_\alpha \fij{33}^{(\alpha)} \\
    i\int_\omega G_{23}^{\rm K} &\equiv f_{23}^{(2)}+\sum_\alpha f_{23}^{(\alpha)}=f_{23} =\frac{1}{2} (\fij{23}^{(2)}+\sum_\alpha Z_\alpha \fij{23}^{(\alpha)})
\end{align}
With the above definition, the equations for the populations in $\ket{2}$ and $\ket{3}$ are easily obtained 
\begin{align}
    &\partial_t n_{22}^{(2)}+i\frac{\Omega}{2}(f_{23}-f_{23}^*)=0 \\
    &\sum_\alpha \frac{1}{Z_\alpha}\partial_t n_{33}^{(\alpha)}-i\frac{\Omega}{2}(f_{23}-f_{23}^*) = 
    \sum_\alpha I_{\rm coll}^{\alpha}.    
\end{align}
From the above equations, the conservation of particles is obtained, i.e., 
\begin{equation}
    \partial_t (N_2+N_3) = \int_{\bp} \partial_t n_{22}^{(2)}+\sum_\alpha 
    \frac{1}{Z_\alpha}\partial_t n_{33}^{(\alpha)} = 0. \label{coherent particle conservation}
\end{equation}
The conservation of particles inside the kinetic equations is a consequence 
of the assumption made on ${\cal A}_{33}$, Eq.~\eqref{spectral 33}, where the 
non-coherent part of the spectral function is totally neglected. 
Finally, in order to close the kinetic equations, only a single polaron 
branch has to be considered. In this case, both $n_{ii}$ 
and $f_{23}$ are projected on the same energy up to an effective mass correction 
and the kinetic equations become 
\begin{align}
    &\partial_t n_2 +i\frac{\Omega}{2}(f_{23}-f_{23}^*) = 0 \label{keq 22 kb} \\
    &\partial_t n_\alpha -iZ_\alpha \frac{\Omega}{2}(f_{23}-f_{23}^*) = I_{\rm coll}^\alpha \label{keq 33 kb} \\
    &\partial_t f_{23} + i\tilde{Z}_\alpha \delta_\alpha f_{23} + i\frac{\Omega}{2}(n_\alpha-n_2) = - \frac{\Gamma_\alpha^{\rm dec}}{2}f_{23}, 
    \label{keq 23 kb}
\end{align}  
with $n_\alpha=n_{33}^{(\alpha)}$, $\delta_\alpha (\bp) = \ve_\alpha(\bp)-\ve_2(\bp)-\Delta$ 
and $\Gamma_\alpha^\mathrm{dec}$ is the decoherence rate. The absence of the tilde accent over collisional integral and decoherence rate indicates that they are written in terms of populations and not of distribution functions.

To better understand the kinetic equations, it is necessary to derive the explicit form of the collisional integral and of the decoherence rate. The strategy is the following: one starts from the definitions, uses the kinetic equation for the dynamics of the molecules, projects on the polaron energy and then write everything in terms of the populations. 

Starting from the definition of the collision integral 
$\tilde{I}_{\rm coll}$ given below Eq.~\eqref{wigner 23}, where Eqs.~(\ref{inst sol 1}-\ref{inst sol 3}) and the definitions of self-energies, Eqs.~(\ref{sig del r}-\ref{skwt}) are used, we obtain the expression
\begin{align}
    \tilde{I}_{\rm coll} &= \left( \frac{1}{2V}\right)^2 \sum_{q,q'} |\gd (q) |^2 {\cal A}_\phi (q-p) {\cal A}_\phi(q-q') {\cal A}_{33}(q') \times \nonumber \\  
    & \times[(F_{33}(p) +F_\phi(q-p))(1+F_\phi(q-q')F_{33}(q')) - (1+F_{33}(p)F_\phi(q-p))(F_{33}(q')+F_\phi(q-q')) ], \nonumber \\    
    \label{I coll 1} 
\end{align}
After the integration over energies, that can be performed using the spectral functions in Eq.~\eqref{I coll 1}, we obtain
\begin{align}
    \tilde{I}_{\rm coll} &= \frac{2\pi}{(2V)^2}\sum_{\bq, \bq'}\sum_{\beta={\rm a,r}}
    |\gd (\omega + \ve_\phi(\bq - \bp),\bq)|^2 Z_\beta(\bq') \delta(\omega + \ve_\phi(\bq - \bp) - \ve_\beta(\bq') - \ve_\phi(\bq - \bq')) \times \nonumber \\
    &\times [(F_{33}(p) +F_\phi(\bq-\bp))(1+F_\phi(\bq-\bq')F_{33}^\beta(\bq')) - (1+F_{33}(p)F_\phi(\bq-\bp))(F_{33}^\beta(\bq')+F_\phi(\bq-\bq')) ].
\end{align}
Note that the full $\gd$ is equivalent to the $T$-matrix, so the notation 
$\gd = T_{\rm sc}$ is used from now on. 
After the on-shell projection, the term inside $\sum_\alpha$ in the 
right-hand side of Eq.~\eqref{kb 33 on-shell F} 
becomes
\begin{align}
    \tilde{I}_{\rm coll}^\alpha &= \frac{2\pi}{(2V)^2}\sum_{\bq, \bq'}\sum_{\beta={\rm a,r}}|T_{\rm sc} (\ve_\alpha(\bp) + \ve_\phi(\bq - \bp),\bq)|^2 Z_\beta(\bq') \delta(\ve_\alpha(\bp) + \ve_\phi(\bq - \bp) - \ve_\beta(\bq') - \ve_\phi(\bq - \bq')) \times \nonumber \\
    &\times [(F_{33}^\alpha(\bp) +F_\phi(\bq-\bp))(1+F_\phi(\bq-\bq')F_{33}^\beta(\bq')) - (1+F_{33}^\alpha(\bp)F_\phi(\bq-\bp))(F_{33}^\beta(\bq')+F_\phi(\bq-\bq')) ].
\end{align}
The physical interpretation of the above is clear: the $T$-matrix plays the role of a cross section, the Dirac delta enforces the energy conservation in the collision and the terms inside the curly brackets are the in and out processes of the collision and $\sum_\beta$ is responsible for both the relaxation and the conversion processes between the polaron branches. Now it is possible to rewrite everything in terms of populations, obtaining the final form of collision integral inside Eq.~\eqref{keq 33 kb}

\begin{equation}
I^\alpha_{\rm coll} = \sum_\beta I_{\alpha \beta},
\label{Icoll final}
\end{equation} 
where 
$\sum_\beta I_{\alpha \beta}$ describes a scattering process between polarons in $\alpha$ and $\beta$. 
$I_{\alpha \beta}$ can be written as 
\begin{equation}
    \label{Icoll final form}
        I_{\alpha\beta}(\bp) = \frac{1}{V} \sum_{\bq'}\left\lbrace
        W_{\bp\bq'}^{\alpha\beta} [1-n_\alpha(\bp)]n_\beta(\bq') 
        - W_{\bq'\bp}^{\beta\alpha} [1-n_\beta(\bp')]n_\alpha(\bq) 
        \right\rbrace.
\end{equation}
In this form, the collisional integral is expressed in terms of the transition rate $W_{\bq'\bp}^{\beta \alpha}$ 
(from the $\alpha$ polaron with momentum $\bp$ to the $\beta$ polaron with momentum $\bq'$) and 
of its complementary process. The transition rates follow the Fermi 
golden rule and are given by 
\begin{align}
    W_{\bp\bq'}^{\alpha\beta} = &\frac{2\pi}{V}\!\! \sum_{\bq}
            |T_\mathrm{sc}( \ve_\beta(\bq')\!+\!\ve_\phi(\bq-\bq'),\bq)|^2
            Z_\alpha(\bq) Z_\beta(\bq') 
            \times \delta( \ve_\alpha(\bp) + \ve_\phi(\bq-\bp) 
                - \ve_\phi(\bq-\bq') - \ve_\beta(\bq')) \nonumber \\
             &\times n_\phi^{\rm eq}(\bq-\bq') 
                [ 1 -  n_\phi^{\rm eq}(\bq-\bp) ],. 
                \label{transition rate}
\end{align} 
{where $n_\phi^{\rm eq}$ is the thermal distribution of the majority atoms with chemical potential $\mu$ and energy $\ve_\phi(\bk) = \bk^2/2m$, $n_\phi^\mathrm{eq}(\bk) = 1/[e^{(\ve_\phi(\bk) - \mu)/T} + 1]$}. 
A very similar procedure to what has been done for the collisional integral can be applied also to the decoherence rate, $\tilde{\Gamma}_{33}^{\text{dec}}=-2\im \sig$, obtaining
\begin{align}
    \tilde{\Gamma}_{33}^{\rm dec}&=\frac{2\pi}{V^2}\sum_{\bq,\bq'}\sum_{\beta={\rm a,r}}|T_{\rm sc}(\ve_\beta(\bq')+\ve_\phi(\bq-\bq'),\bq)|^2Z_\beta(\bq')\delta(\omega + \ve_\phi(\bq-\bp)-\ve_\phi(\bq-\bq')-\ve_\beta(\bq')) \nonumber \\ 
    &\times [(1-n_\phi^{\rm eq}(\bq-\bq'))(1-n_\beta(\bq'))n_\phi^{\rm eq}(\bq-\bp) +(1-n_\phi^{\rm eq}(\bq-\bp))n_\phi^{\rm eq}(\bq-\bq')n_\beta(\bq') ].
\end{align}
After the projection on polaron energies and the substitution of distribution functions 
with population it is straightforward to obtain the final form of the 
decoherence rate 
\begin{equation}
    \Gamma_\alpha^{\rm dec}=\frac{\tilde{Z}_\alpha(\bp)}{Z_\alpha(\bp)}
    \frac{1}{V}\sum_{\bq'}\sum_{\beta={\rm a,r}}W_{\bp \bq'}^{\alpha \beta}
    n_\beta(\bq')+W_{\bq'\bp}^{\beta \alpha}(1-n_\beta(\bq')),
    \label{deco rate final}
\end{equation}
where the transition rates defined in Eq.~\eqref{transition rate} have been used. 
In this expression, the first term is related to the in processes, while the second is related 
to the out ones. Note that while both of them are originating from 
collisions of the impurity atom with the bath, collisional integral 
and decoherence rate are different in nature. The first one is indeed 
related to the imbalance of populations at different momenta and 
eventually at different branch, the second one affects coherences. 
In the long time, the collisional integral drives the population 
of minority atoms in thermal equilibrium with the bath. 

{In the impurity limit, where $n_\alpha(\bk) \approx 0$, the decoherence rate takes the form
\begin{equation}
    \Gamma^{\mathrm{dec}}_{\alpha}(\bk) \!=\!
    \frac{\tilde Z_\alpha(\bk)}{Z_\alpha(\bk)}\! \frac{1}{V}\!\!
    \sum_{\bk',\beta}W_{\bk'\bk}^{\beta\alpha},
\end{equation}
while $I_{\alpha \beta}$ can now be written as 
\begin{equation}
    I_{\alpha\beta}(\bk) = \frac{1}{V} \sum_{\bk'}\bigg[
    W_{\bk\bk'}^{\alpha\beta} n_\beta(\bk') 
    - W_{\bk'\bk}^{\beta\alpha} n_\alpha(\bk)
    \bigg].
\end{equation}
}

As briefly mentioned above, it is important to notice that, due to the omission of the back-flow term, the kinetic equations Eqs.~(\ref{keq 22 kb}-\ref{keq 23 kb}) conserve the sum
$N_3(t) + \tilde N_2(t)$, with
$\tilde N_2(t) = \sum_\bk Z_\alpha n_2(\bk,t)$ and $N_3(t) = \sum_{\beta,\mathbf{k}}n_\beta(t,\mathbf{k})$, where the polaron $\alpha$ is on resonance with the non-interacting state. Since at the initial time, a fraction $1-Z_\alpha$ is missing from $\tilde N_2$ with respect to the total $N_2$, in order to implement the conservation law we assume for consistency that this missing part is not evolving in time and directly add it to $\tilde N_2(t)$ to obtain $N_2(t)$. Therefore, the experimental observable $N_2/(N_2+N_3)$ can be rewritten as $1-{\sum_{\alpha,\mathbf{k}}}n_\alpha(t,\mathbf{k})/N_2(t=0)$. We note that the same form can be obtained by using the total number conservation to rewrite $N_2(t)/(N_2(t)+N_3(t))=1-N_3(t)/(N_2(t)+N_3(t))$. By adding to kinetic equations back-flow terms and gradient corrections to the collision integral encoding memory effects, proper particle-number conservation can be achieved \cite{Knoll2001}, but at the cost of largely increasing the complexity of the equations as well as the computational demand of their solution.
This is beyond the scope of the present work and will be the subject of future investigations.

\section{Repulsive polaron}
{In this section, we provide a comparison between the observed Rabi oscillation for the repulsive polaron, see Ref.~\cite{scazza2017repulsive}, and the theoretical predictions derived using the kinetic equations, Eqs.~(\ref{keq 22 kb}-\ref{keq 23 kb}). To do so, we have to compute $\Gamma_\rep^{\mathrm{dec}}$ and $I_\rep$ and insert them into kinetic equations. For the decoherence rate, we cannot simply replace} a$\rightarrow$r in Eq.~(7) of the main text, because $\Gamma_\rep^{\mathrm{dec}}$ and $I_\rep$ also contain terms $\propto W^{\att\rep}$, describing collisions that result in de-excitation to the attractive branch. However, the observed decay rate~\cite{scazza2017repulsive} suggests that such a process is absent at weak interactions, where it is instead replaced by a much slower three-body recombination to the molecular state. {Since we will present results for $k_F a >0$ and far from unitarity, we will not consider the de-excitation process.
}

\begin{figure}[t]
  \centering
  \includegraphics[width=0.75\columnwidth]{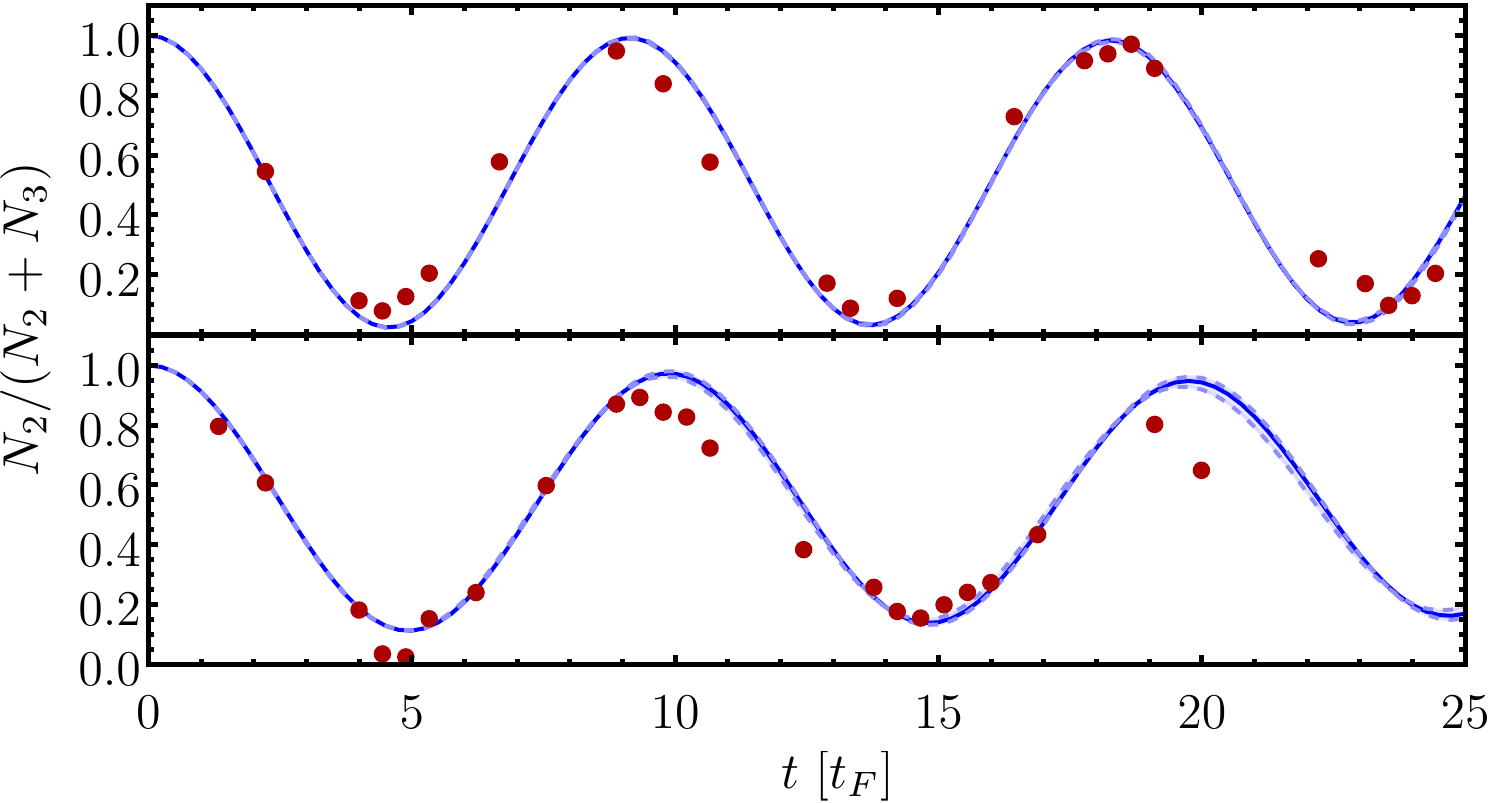}
\caption{Rabi oscillations of the repulsive polarons for $1/(k_Fa)=$ 2.63 and 1.27 (upper and lower panels, respectively). The blue solid line is calculated from Eq.~\eqref{eq:keq}, whereas the red dots are from the experiment~\cite{scazza2017repulsive}. The parameters are $\Omega=0.7\ve_F$, $T=0.135\ve_F$, $\delta_\rep(\bk=0)=0$. The shaded blue region (mostly unnoticeable) corresponds to a $20\%$ uncertainty in the temperature of the majority component.}
\label{fig_rep}
\end{figure}

Our predictions for the Rabi oscillations of repulsive polarons, i.e., the case when the laser drive is resonant with the transition between state $|2\rangle$ and the repulsive polaron branch, i.e., $\delta_\rep(0)=0$,
are reported in Fig.~\ref{fig_rep}, {where we plot the same quantity of the main text, i.e.,
\begin{equation}
\frac{N_2(t)}{N_2(t) + N_3(t)} = 1-\frac{\sum_{\alpha,\bk}n_\alpha(t,\bk)}{N_2(t=0)}.
\end{equation}
}
We consider the experimentally relevant cases $1/(k_Fa) = 2.63$ and 1.27, corresponding to the polaron parameters $(E_\rep/\varepsilon_F,\,  m_*/m,\,  Z_\rep(0))$ = $(0.442,\, 1.13,\, 0.823)$ and $(0.188,\, 1.07,\, 0.963)$, respectively, {see Tab.~\ref{tab:Repulsive polaron}. 
}


The dots correspond to the experimental results from Ref.~\cite{scazza2017repulsive}. To simplify the problem, as before, we assumed that no population is coherently transferred to the attractive branch, i.e., $n_\att(t,{\bf k})=0$. 
 Neglecting the decay to the attractive branch, the decoherence rate of the repulsive polarons $\Gamma_\rep^\mathrm{dec}$ is not noticeable on the timescale considered. 
In Fig.~\ref{fig_rep}, we observe indeed coherent Rabi oscillations given by the renormalized frequency $\sqrt{Z_\rep}\Omega$  
without any significant decay of the signal, in good agreement with the experimental data.
 For these parameter regimes, the uncertainty in the temperature (shaded region) is almost unnoticeable.

For the parameters considered here, the collision integral plays only a minor part, and neglecting it is a very good approximation for the observables shown in Fig.~\ref{fig_rep}, which are dominated by the coherent Rabi transfer between the states and the decoherence rate $\Gamma_\alpha^\mathrm{dec}$. 
The smallness of the collision integral also implies slow thermalization of the impurity. 

Interestingly, as pointed out in the experiment \cite{scazza2017repulsive}, it appears that the theoretically predicted strong decay of the repulsive polaron to the attractive branch does not play a role in the Rabi dynamics (see also the discussion in \cite{haydn2020repulsive}).

\begin{figure}[t]
  \centering
  \includegraphics[width=0.9\columnwidth]{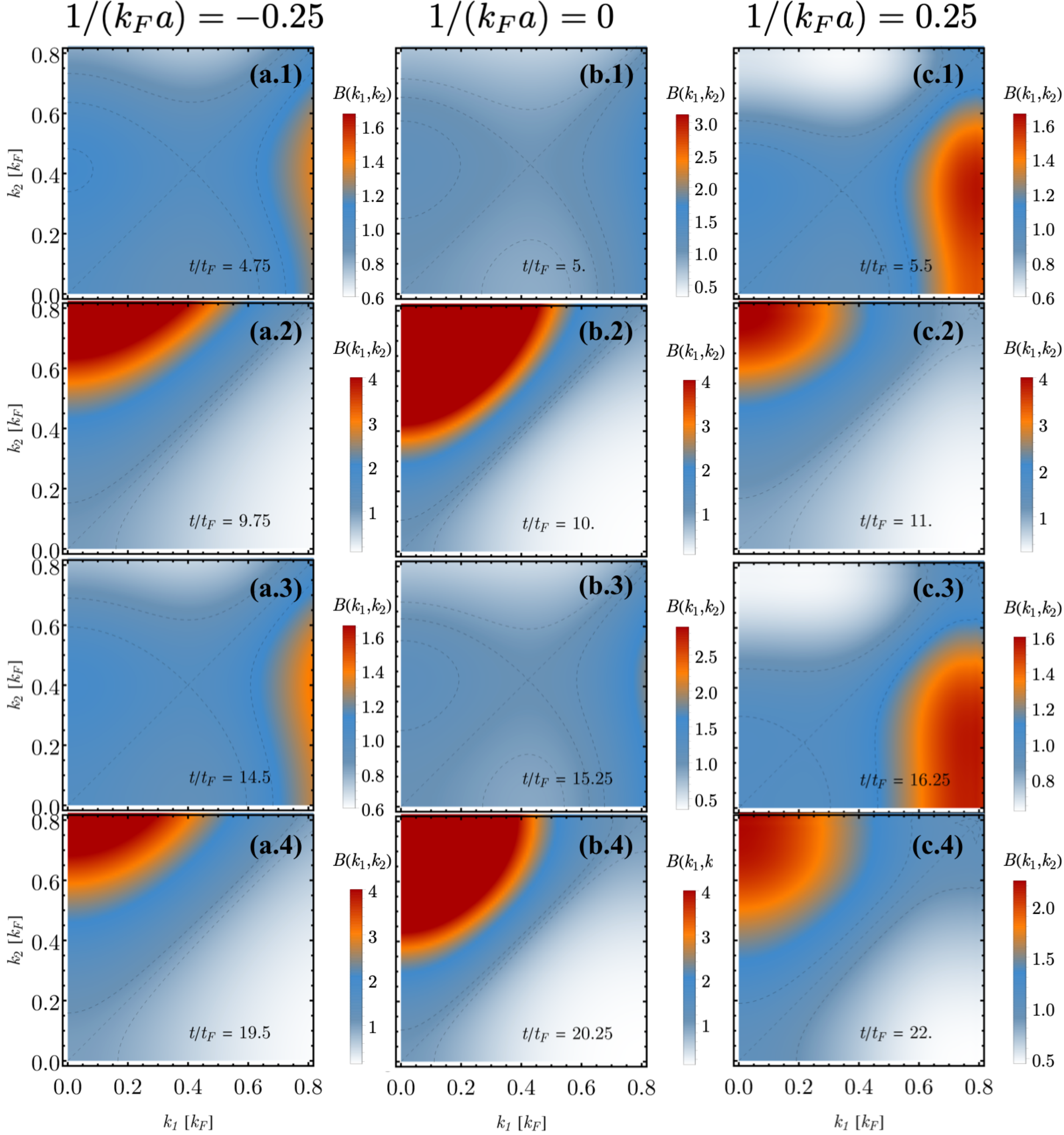}
\caption{The map of the violation of the detailed balance quantified by the parameter $B(k_1,k_2)$ from Eq.~\eqref{Bff}. The dashed lines mark the contours where $B= 0.9$, 1, 1.1; notice that $B=1$ also on the diagonal $k_1=k_2$. The interaction strengh is $1/(k_Fa) = -0.25$ (left column), 0 (middle column), +0.25( right column). Each row corresponds to a minimum (the first and third rows) or maximum (the second and fourth rows) of the Rabi oscillations shown in Fig.~2 of the main text.
The parameters: $\Omega=0.7\ve_F$, $T=0.135\ve_F$, $\delta_\att(0)=0$.}
\label{fig_dbv}
\end{figure}

\section{Thermalization and bound on the violation of detailed balance}

In this Section we provide additional results for the time-dependent violation of the detailed balance 
in the approach of the polaron to a thermal state. We also derive a bound on the average detailed balance violation 
given by the Kullback-Leibler divergence as mentioned in the main text. Finally, we also discuss the shape of the distributions of the polarons.

\subsection{{Detailed balance violation}}

The detailed balance in the main text is probed by the following quantity:
\begin{equation}
    B \equiv \frac{1 - n_\att(\bk)}{n_\att(\bk)}
            \frac{1 - n_1(\bk_1)}{n_1(\bk_1)}
            \frac{n_\att(\bk')}{1 - n_\att(\bk')}
            \frac{n_1(\bk_1')}{1 - n_1(\bk_1')}.
\end{equation}
If the environment $n_1$ is in equilibrium, this object simplifies to the following form, which depends only on two momenta:
\begin{equation}
    B(\bk,\bk') = \frac{f(\bk)}{f(\bk')}, 
\label{Bff}
\end{equation}
where, defining $\beta = 1/T$, 
\begin{equation}
    f(\bk) = \frac{1 - n_\att(\bk)}{n_\att(\bk)} e^{-\beta \ve_\att(\bk)}.
\label{Bsimp}
\end{equation}
In this notation, we dropped the time dependence of $n_\att$ for brevity. If the polarons are in a thermal state, $B(\bk,\bk')\equiv 1$ for all momenta, and the deviation from 1 signals the violation of detailed balance. If, on the other hand, $B(\bk,\bk')>1$, the rate of the \emph{in} scattering processes, which populate the mode $\bk$ as a result of collisions of polarons with momentum $\bk'$, is larger than the \emph{out} processes, in which polarons with momentum $\bk$ are scattered out to momentum $\bk'$.

In Fig.~\ref{fig_dbv}, we plot $B(\bk_1, \bk_2)$ for times corresponding to extrema of Fig. 2 in the main text for $1/(k_F a) =$ -0.25 (left column), 0 (middle column), 0.25 (right column). The dashed lines mark the values 0.9, 1.0 (notice also $B(k_1,k_1) = 1$ on the diagonal), 1.1. Here, we observe that, for the times corresponding to the minima of Rabi oscillations, when the polaron population is maximal, the region in momentum space (see the first and third rows), where the detailed balance is $B\approx 1$, is large. This is also manifested as minima of the main panel in Fig. 3 in the main text. On the contrary, a large region with significant violation of the detailed balance is seen for times when the population is very small (see the second and fourth rows), i.e., at the maxima of the oscillations.

From Fig. 4, in the second and fourth rows, which correspond to minima of the polaron densities in the Rabi oscillations, a simple structure forms. Namely, for all $k_2<k_1$ we have $B(k_1,k_2) < 1$, which means that the \emph{out} scattering from $k_1$ to $k_2<k_1$ overcomes the \emph{in} collisions, and we observe an effective cooling of the system. This effect is also visible in the inset of Fig. 3 of the main text, where the inferred temperature $T_K$ decreases at times corresponding to maxima of the Rabi oscillations (see Fig. 2 in the main text).

If the system is close to thermal equilibrium, $B \approx 1$ and the deviation $B-1$ quantifies the degree of detail balance violation. In the main text, we average this deviation over the system, to obtain a single measure of the violation. Thus, we have
\begin{equation}
    (\Delta B)^2 \equiv \langle [B(\bk,\bk') - 1 ]^2 \rangle,
\end{equation}
where the average is taken over the normalized distribution $\propto n_\att(\bk) n_\att(\bk')$. Since Eq.~\eqref{Bff} is a product of two factors, we obtain:
\begin{equation}\label{DB2}
    (\Delta B)^2 = 1 + \langle f(\bk)^2 \rangle \langle f(\bk)^{-2} \rangle - 2 \langle f(\bk) \rangle \langle f(\bk)^{-1} \rangle,
\end{equation}
where $\langle \cdot \rangle \equiv C \sum_\bk (\cdot  n_\att(\bk))$, and $C$ is the normalization constant.

Now, we assume that $n_\att(\bk) = n_T(\bk) + \delta n_\bk$, where $n_T = 1/(e^{\beta (\ve_\att(\bk)- \mu_a)} +1)$ is a thermal state, and the chemical potential $\mu_a$ is adjusted in order to match the density of the polarons.

Expanding now Eq.~\eqref{DB2} to second order in $\delta n_\bk$, we obtain
\begin{eqnarray}
    \langle f(\bk)^2 \rangle &=& \langle  1 + 2 f_1(\bk) + (f_1^2(\bk) + 2 f_2(\bk))   \rangle,\\
    \langle f(\bk)^{-2} \rangle &=& \langle  1 - 2 f_1(\bk) + (3f_1^2(\bk) - 2 f_2(\bk))  \rangle,\\
    \langle f(\bk) \rangle &=& \langle  1 + f_1(\bk) + f_2(\bk)  \rangle,\\
    \langle f(\bk)^{-1} \rangle &=& \langle  1 - f_1(\bk) + (f_1^2(\bk) - f_2(\bk))  \rangle,
\end{eqnarray}
where $f_1(\bk) = - (\delta n_\bk e^{-\beta \ve_\att(\bk)}) / n_T^2(\bk)$ and $f_2(\bk) = (\delta n_\bk^2 e^{-\beta \ve_\att(\bk)}) / n_T^3(\bk)$

Up to the second order in $\delta n_\bk$ we obtain
\begin{equation}
    (\Delta B)^2 = 2 \big[ \langle f_1^2 \rangle - \langle f_1 \rangle^2 \big] \geqslant 0.
\end{equation}
Since we can drop the second term in order to get the bound $\Delta B^2 \leqslant 2 \langle f_1^2 \rangle$, we have
\begin{equation}
    \langle f_1^2 \rangle = C \sum_\bk n_\att(\bk)\bigg( - \frac{\delta n_\bk e^{-\beta \ve_\att(\bk)}}{n_T^2(\bk)} \bigg)^2 
    \approx C \sum_\bk \frac{\delta n_\bk^2}{[1-n_T(\bk)]^2n_T(\bk)},
\end{equation}
where we replaced $n_\att$ by $n_T$, since we are concerned with the second order in $\delta n_\bk$.

Now, we use $1/[n_T(1-n_T)^2] = (1 + e^{-\beta \ve_\att})^2 / n_T$, and we obtain:
\begin{equation}
    \langle f_1^2 \rangle = C \sum_\bk \frac{\delta n_\bk^2}{n_T(\bk)} \bigg( 1 + e^{-\beta \ve_\att(\bk)} \bigg)^2
    \leqslant \bigg( 1 + e^{-\beta \ve_\att(0)} \bigg)^2 C \sum_\bk \frac{\delta n_\bk^2}{n_T(\bk)}.
\end{equation}

Finally, since up to the second order, the Kullback-Leibler divergnce has the form $D(\tilde n_\att | \tilde n_T) \approx C \sum_\bk \delta n_\bk^2 / (2n_T(\bk))$, here tilde indicates the normalization, we obtain
\begin{equation}
    (\Delta B)^2 \leqslant 4 \bigg( 1 + e^{-\beta \ve_\att(0)} \bigg)^2 D(\tilde n_\att | \tilde n_T),
\end{equation}
which has the form mentioned in the main text, i.e.,  $\langle (B-1)^2\rangle \leqslant \zeta D(\tilde n_\att | \tilde n_T)$ with 
$\zeta = 4(1 + e^{-\beta \ve_\att(0)})^2 > 0$.

\begin{figure}[!t]
  \centering
  \includegraphics[width=0.4\columnwidth]{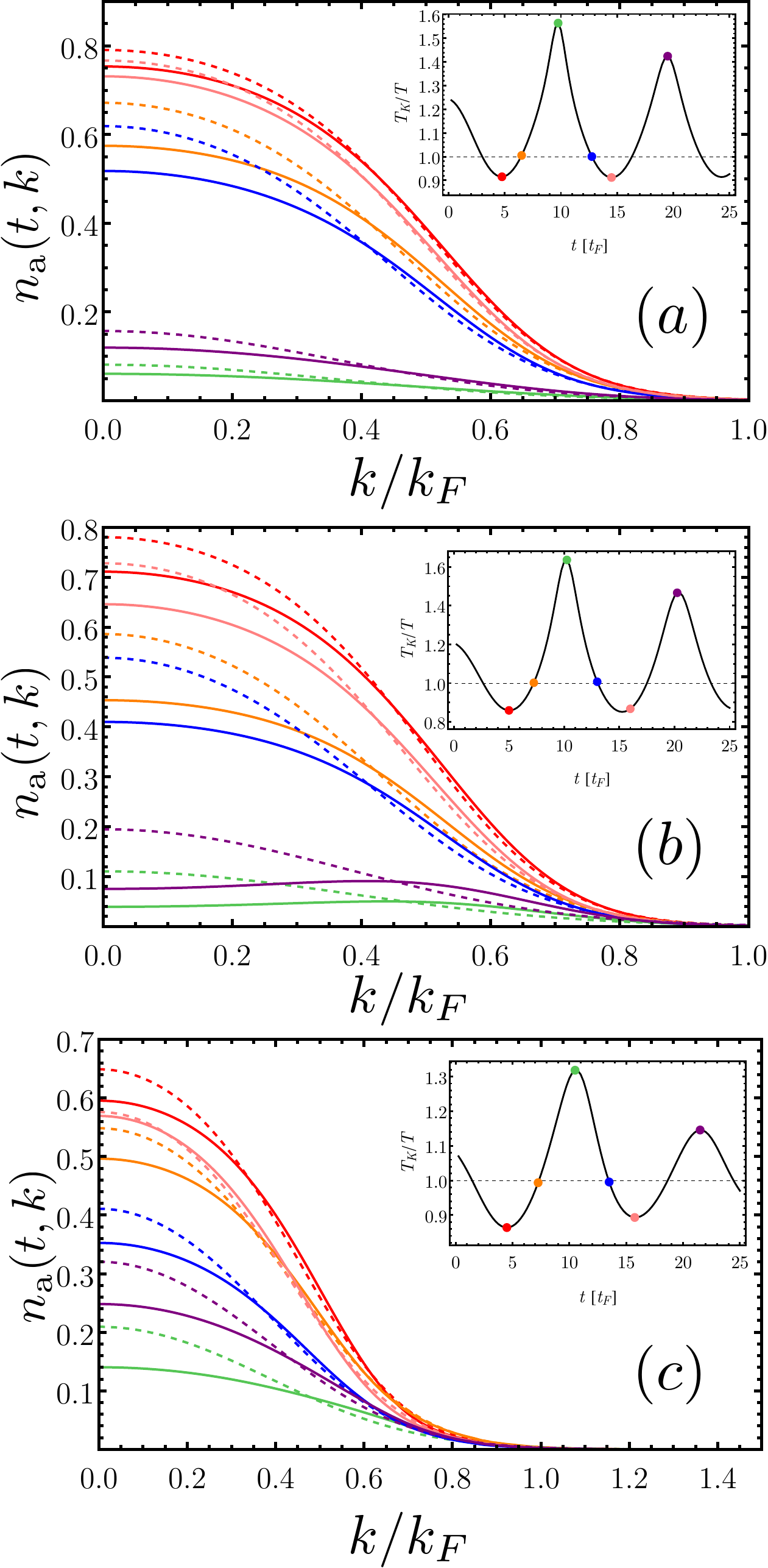}
\caption{The numerically determined occupations of $n_\mathrm{a}(t,\bk)$ (solid) and the inferred equilibrium distribution $n_\att^\mathrm{eq}(T_K(t)),\bk)$ (dashed) as a function of momentum $k=|\bk|$ (in the units of $k_F$) for various times corresponding to the points indicated in the insets. The color code of the curves in the main panel is the same as the times corresponding to the colors of the points in the inset. Main panel: from top to bottom: $1/(k_F a) = $ -0.25 (panel a), 0 (b), 0.25 (c). The inset: shows the inferred temperature $T_K$ as a function of time $t$ (in units of $t_F$) as described in the main text.
The parameters: $\Omega=0.7\ve_F$, $T=0.135\ve_F$.
}\label{fig_distr}
\end{figure}

\subsection{{Occupation of attractive polarons}}

In Fig.~\ref{fig_distr}, we show the occupations of the momenta of the attractive polaron branch, i.e., $n_\att(t,\bk)$, for $1/(k_F a) =$ -0.25 (panel a), 0 (b), 0.25 (c) at some characteristic times $t$. We have chosen the times $t$ corresponding to the minima (red and pink points in the insets), $T_K/T = 1$ (orange and blue) and maxima (green and purple) of the inferred temperature $T_K$ which varies with time as shown in the corresponding insets. The solid curves show the results of the numerical simulations of $n_\att(t,\bk)$ as a function of $k=|\bk|$ while the dashed curves are the optimal thermal distribution functions $n_\att^\mathrm{eq}(T_K,\bk)$ corresponding to the optimized temperatures $T_K$.

The presented results show that the overall shape of the distributions $n_\att(t,\bk)$ is captured by the found equilibrium distribution, which explains the fact that the inferred $T_K$ is of the correct order as $T$. All the nearest equilibrium distributions are, however, overestimated at low momenta [apart from the dip in the distribution for the green curve in (b) corresponding to the first maximum of $T_K/T$]. This low momentum region, however, is weakly represented in the Kullback-Leibler divergence due to the jacobian $k^2$ in the integration. The worst agreement is found for low densities of the polarons, which correspond to the peaks of $T_K$ (green and purple points in the insets), since the distribution is very far from equilibrium. Here, we note that the non-Boltzmann-like shape of the distributions confirms the need for the non-equilibrium theory of the Rabi-coupled dynamics of polarons.

Finally, we comment on the timescales of the approach to the steady state. Since it is challenging to extract its form analytically due to non-trivial dependence of the $T$-matrix on momentum $\bk$, we resorted to its simple estimation from the dynamics of the violation of detailed balance. Assuming an exponential decay, we fitted an exponential function $\propto e^{-t/\tau_\mathrm{th}}$ to the peaks that are shown in Fig.~3 in the main text. We have found 
$\tau_\mathrm{th}=$ 
22 $t_F$ for $1/(k_F a) = $ -0.25, 
29 $t_F$ for $1/(k_F a)$ = 0 and 
14 $t_F$ for $1/(k_F a)$ = +0.25.
The markedly shorter time for $1/(k_F a)$ = +0.25 captures the behaviour of the red dotted curve from Fig. 3 (main text, main panel) which approaches zero on a faster timescale. Interestingly, these simple results show non-monotonic behaviour of thermalization across the unitary limit $1/(k_F a)=0$, which is also visible in Fig. 3 of the main text.